\documentclass[twocolumn,showpacs,pra]{revtex4-2}
\usepackage[latin9]{inputenc}
\usepackage{amsmath}
\usepackage{amssymb}
\usepackage{esint}
\usepackage{graphicx}
\usepackage{hyperref}

\usepackage{physics}
\usepackage{soul}
\usepackage[makeroom]{cancel}
\usepackage{placeins}
\usepackage[normalem]{ulem} 

\makeatletter
 \@ifundefined{textcolor}{}
 {%
   \definecolor{BLACK}{gray}{0}
   \definecolor{WHITE}{gray}{1}
   \definecolor{RED}{rgb}{1,0,0}
   \definecolor{GREEN}{rgb}{0,1,0}
   \definecolor{BLUE}{rgb}{0,0,1}
   \definecolor{CYAN}{cmyk}{1,0,0,0}
   \definecolor{MAGENTA}{cmyk}{0,1,0,0}
   \definecolor{YELLOW}{cmyk}{0,0,1,0}
 }

\usepackage{dcolumn}\usepackage{bm}\usepackage{color}

\makeatother

\begin{document}

\title{Phase-space stochastic quantum hydrodynamics for interacting Bose gases}
\author{S. A. Simmons}
\affiliation{School of Mathematics and Physics, The University of Queensland, Brisbane, Queensland 4072, Australia}
\author{J. C. Pillay}
\affiliation{School of Mathematics and Physics, The University of Queensland, Brisbane, Queensland 4072, Australia}
\author{K. V. Kheruntsyan}
\affiliation{School of Mathematics and Physics, The University of Queensland, Brisbane, Queensland 4072, Australia}

\date{\today}

\begin{abstract}

Hydrodynamic theories offer successful approaches that are capable of simulating the otherwise difficult-to-compute dynamics of quantum many-body systems. In this work we derive, within the positive-$P$ phase-space formalism, a new stochastic hydrodynamic method for the description of interacting Bose gases. It goes beyond existing hydrodynamic approaches, such as superfluid hydrodynamics or generalized hydrodynamics, in its capacity to simulate the full quantum dynamics of these systems: it possesses the ability to compute non-equilibrium quantum correlations, even for short-wavelength phenomena. Using this description, we derive a linearized stochastic hydrodynamic scheme which is able to simulate such non-equilibrium situations for longer times than the full positive-$P$ approach, at the expense of approximating the treatment of quantum fluctuations, and show that this linearized scheme can be directly connected with existing Bogoliubov approaches. Furthermore, we go on to demonstrate the usefulness and advantages of this formalism by exploring the correlations that arise in a quantum shock wave scenario and comparing its predictions to other established quantum many-body approaches.
\end{abstract}
\maketitle

\section{Introduction}

Understanding and simulating the far-from-equilibrium behavior of quantum many-body systems is a notoriously challenging task. A particularly powerful method capable of describing such non-equilibrium behavior is the incredibly successful hydrodynamic approach. Here our interest lies in the application of hydrodynamic theories to the out-of-equilibrium dynamics of ultra-cold quantum fluids, where there has been significant interest in recent years, predominantly owing to the development of the theory of generalized hydrodynamics (GHD) \cite{Yoshimura2016,Fagotti2016,Doyon2017,Dubail2019,Weiss2021,Bouchoule2022}, as well as the unforeseen success of existing theories of classical hydrodynamics to situations where their \textit{a priori} applicability would have been questionable \cite{Bouchoule2016,Atas2019}. Since hydrodynamic theories are often postulated phenomenologically, we wish to gain a deeper understanding of their emergence from a microscopic description of the system, and to provide such a microscopic theory in terms of hydrodynamic variables that is capable of going beyond existing hydrodynamic approaches to include quantum correlations and short-wavelength phenomena.

Conventional or classical hydrodynamics (CHD) itself is postulated using the local density approximation (LDA) where the fluid is treated by breaking it up into small uniform slices which are then assumed to be in thermal equilibrium in the local moving frame. Nevertheless, the classical hydrodynamic equations themselves are in fact nothing more than equations which express the conservation of mass, momentum, and energy (or entropy) of a system \cite{Euler1757,Landau&LifshitzV6}. Hence, their applicability goes beyond that of a \textit{classical} ideal gas and they can be used to simulate the large scale dynamics of both weakly and strongly interacting \textit{quantum} fluids by employing the appropriate thermodynamic equation of state \cite{Bouchoule2016,Atas2019,Minguzzi2001,Menotti2002,Peotta2014,Abanov2012}. Yet such an approach is limited in its applicability in that it usually requires that a system possess fast themalization times, and due to the LDA it is restricted to the description of long-wavelength excitations.

To go beyond the classical hydrodynamic description, one can employ superfluid hydrodynamics, often referred to as quantum hydrodynamics, which is a well-known and fundamental tool used to describe ultra-cold interacting Bose gases \cite{pitaevskii_stringari_2016,Pethick&Smith2008}. Through the use of Madelung's transformation \cite{Madelung1927}, it is usually derived from the mean-field Gross-Pitaevskii equation (which describes weakly interacting Bose-Einstein condensates of ultra-cold atomic gases at zero-temperature) and goes beyond the classical hydrodynamic approach to include the so-called quantum pressure term, which allows for the accurate description of short-wavelength phenomena.
It is also possible to extend such a description to finite temperatures by employing a two-fluid model where the single fluid is separated into two components; the condensed atoms (or the superfluid component) which are in the ground-state of the system, and the non-condensed atoms that comprise the thermal cloud (or the normal component) \cite{Griffin2000}. Here, the condensed atoms are treated according to superfluid hydrodynamics and are then coupled to the non-condensed atoms that are treated classically. Such an approach can be directly connected to Landau's celebrated two-fluid hydrodynamic formalism, which itself was postulated phenomenologically to describe the strongly interacting system of superfluid Helium \cite{Tisza1938,Landau1941,pitaevskii_stringari_2016}. Despite the powerful utility of the superfluid hydrodynamic approach, it relies fundamentally on the assumption of a mean-field or existence of long-range order, which neglects any quantum fluctuations and correlations, and as such, it is not a truly quantum description.

The theory of GHD exceeds that of classical hydrodynamics in a different manner to superfluid hydrodynamics, and it proceeds through the identification of infinitely many conserved quantities that exist for integrable systems. Including these higher order conserved quantities in its description allows GHD to simulate scenarios where CHD would otherwise suffer from the notorious gradient catastrophe (or derivative discontinuity) problem, and provides a more accurate description at finite-temperatures \cite{Doyon2017}. Standard GHD is capable of simulating ultra-cold Bose gases at any interaction strength, and since its discovery it has been extended to address (among other scenarios) near-integrable systems \cite{Yoshimura2017,Bouchoule2022} and to include quantum fluctuations \cite{Jerome2020,Ruggiero2021}. Yet, to date, these descriptions remain restricted to long-wavelength excitations and are unable to capture quantum interference and other important short-wavelength phenomena such as those present in dispersive quantum shock waves \cite{Hoefer2016,Dutton2001,Damski2004,Simula2005,Cornell2006,Hoefer2008,Davis2009,Abanov2012,Bulgac2012,Peotta2014,Simmons2020}.

It is an interesting question to ask then, whether an exact quantum description of interacting Bose gases (which is capable of capturing both quantum correlations and short-wavelength phenomena) is possible in terms of hydrodynamic variables and what it would look like. In this work we show that it is possible, in the weakly-interacting regime, and that such a theory can be formulated in a phase-space representation; we refer to it as \textit{stochastic quantum hydrodynamics} (SQHD). By employing the appropriate assumptions (of small quantum fluctuations), we additionally derive the new approximate hydrodynamic equations of \textit{linearized stochastic quantum hydrodynamics} (LSQHD) and provide connections between these and other existing approaches capable of treating interacting Bose gases. Furthermore, we apply our novel approach to the dynamics of dispersive quantum shock waves \cite{Simmons2020} where we calculate for the first time the quantum correlations arising in this situation and benchmark our results against alternate theoretical approaches.

\section{Exact quantum hydrodynamics: stochastic formulation}

We consider an effective field theory Hamiltonian for a trapped Bose
gas with repulsive delta-function interactions in second-quantized form
\begin{equation}
	\hat{H}=\int d {\mathbf{r}}\left\{ \frac{-\hbar^{2}}{2m}\hat{\Psi}^{\dagger}\laplacian\hat{\Psi}+V_{\mathrm{ext}}( {\mathbf{r}},t)\hat{\Psi}^{\dag}\hat{\Psi}+\frac{g}{2}\hat{\Psi}^{\dag}\hat{\Psi}^{\dag}\hat{\Psi}\hat{\Psi}\right\} , \label{H}
\end{equation}
where $\hat{\Psi}( {\mathbf{r}},t)$ is the bosonic field operator with the usual equal-time commutation relation $[\hat{\Psi}( {\mathbf{r}},t),\hat{\Psi}^{\dagger}( {\mathbf{r}}^{\prime},t)]=\delta^{(3)}( {\mathbf{r}}- {\mathbf{r}}^{\prime})$ and $m$ is the atomic mass.
The first term in Eq.~\eqref{H} is the kinetic energy, and the last term describes the $s$-wave scattering interactions between the particles, with the coupling strength $g=4\pi\hbar^{2}a/m>0$ in three dimensions (3D), where $a$ is the $s$-wave scattering length. The external trapping potential is given by $V_{\mathrm{ext}}( {\mathbf{r}},t)$, which can in general be time-dependent. 

In an elongated (cigar-shaped) trap geometry, this model can easily be reduced to the trapped gas Lieb-Liniger model \cite{LiebLiniger1963} in one-dimension (1D) if $a$ is much smaller than the amplitude of transverse zero point oscillations (or the harmonic oscillator length) $l_{\perp}=\sqrt{\hbar/m\omega_{\perp}}$, where $\omega_{\perp}$ is the frequency of transverse (radial) confinement assumed harmonic and symmetric. In this case, the interaction constant is  given by $g\rightarrow g_{\mathrm{1D}}\simeq 2\hbar\omega_{\perp}a$ away from confinement induced resonances \cite{Olshanii98}. Similarly, in a pancake geometry, the model can be reduced to two dimensions (2D), with $g\rightarrow g_{\mathrm{2D}}=2\sqrt{2\pi}\hbar^2a/ml_0$ \cite{petrov2004low}, where $l_0$ is the harmonic oscillator length in the axial dimension.

%We further note that the use of the effective delta-function interaction potential $g\delta( {\mathbf{r}}- {\mathbf{r}}^{\prime})$ between atom pairs, leading to Eq. (\ref{H}), must assume a UV momentum cutoff $k^{\max}$ to prevent the theory from UV divergences \cite{abrikosov1965quantum}. In the numerical simulations below, such a momentum cutoff is imposed explicitly via the finite computational lattice. If the lattice spacings ($\Delta x$, $\Delta y$, $\Delta z$) in each spatial dimension are chosen to be much larger than the $s$-wave scattering length $a$, then the respective momentum cutoffs satisfy $k_{x,y,z}^{\max}\ll1/a$. In this case, the coupling constant $g$ can be approximated by the expression given above, without the need for explicit renormalization \Rev{\cite{Morgan_2000,Castin2003}}.

We further note that use of the contact interaction potential $g\delta( {\mathbf{r}}- {\mathbf{r}}^{\prime})$ in the Hamiltonian  (\ref{H}), instead of the true interatomic potential with non-zero range, is restricted to low energies and momenta. Accordingly, it must be used together with a UV momentum cutoff as to prevent the theory from ultraviolet divergences that show up in perturbative schemes \cite{abrikosov1965quantum,Pethick&Smith2008,pitaevskii_stringari_2016,Morgan_2000}. % (Alternatively, such divergencies can be removed by the procedure of remormalization and the use of the $T$-matrix approach.) 
In the numerical simulations below, the required UV momentum cutoff is imposed explicitly via the finite computational lattice. If the lattice spacings ($\Delta x$, $\Delta y$, $\Delta z$) in each spatial dimension are chosen to be much larger than the $s$-wave scattering length $a$, then the respective momentum cutoffs satisfy $k_{x,y,z}^{\max}\ll1/a$ and the coupling constant $g$ in the contact potential can indeed be expressed in terms of the 3D scattering length $a$ as $g=4\pi\hbar^{2}a/m$ \cite{Morgan_2000,Sinatra2002,Castin2003}. Physically, the restriction $k_{x,y,z}^{\max}\ll1/a$ means that the theory with contact interactions cannot be used to describe phenomena on lengthscales shorter $a$.

To model the dynamics of quantum fields governed by the above Hamiltonian, we use the positive-$P$ phase-space approach \cite{Drummond1980,GardinerCrispin1985,Steel1998,Hope2001,Deuar2006,Deuar2007,Perrin2008,Kheruntsyan2010}. In this approach, the dynamics of the bosonic field operators $\hat{\Psi}(\mathbf{r},t)$ and $\hat{\Psi}^{\dagger}(\mathbf{r},t)$ are equivalent to the evolution of two independent complex $c$-fields, $\Psi(\mathbf{r},t)$ and $\widetilde{\Psi}(\mathbf{r},t)$, satisfying the \^{I}to stochastic differential equations (SDEs) \cite{Deuar2007,Perrin2008,Kheruntsyan2010}: 
\begin{eqnarray}
	\frac{\partial\Psi(\mathbf{r},t)}{\partial t} & = & \frac{i}{\hbar}\left(\frac{\hbar^{2}}{2m}\laplacian-V_{\mathrm{ext}}(\mathbf{r},t)-g\widetilde{\Psi}\Psi\right)\Psi\nonumber \\
	 &  & +\sqrt{-i\frac{g}{\hbar}}\Psi\,\xi_{1}(\mathbf{r},t),\label{eq:pos-P-1}\\
	\frac{\partial\widetilde{\Psi}(\mathbf{r},t)}{\partial t} & = & -\frac{i}{\hbar}\left(\frac{\hbar^{2}}{2m}\laplacian-V_{\mathrm{ext}}(\mathbf{r},t)-g\widetilde{\Psi}\Psi\right)\widetilde{\Psi}\nonumber \\
 	&  & +\sqrt{i\frac{g}{\hbar}}\widetilde{\Psi}\,\xi_{2}(\mathbf{r},t),\label{eq:pos-P-2}
\end{eqnarray}
where $\laplacian=\boldsymbol{\nabla}\cdot\boldsymbol{\nabla}$ is the Laplacian operator, $\xi_{j}(\mathbf{r},t)$ ($j=1,2$) are real independent Gaussian noise sources with zero mean and the following nonzero correlations $\langle\xi_{j}(\mathbf{r},t)\xi_{k}(\mathbf{r}^{\prime},t^{\prime})\rangle=\delta_{jk}\delta^{(3)}(\mathbf{r}-\mathbf{r}^{\prime})\delta(t-t^{\prime})$.

The above SDEs are derived from an equivalent Fokker-Planck equation for the positive-$P$ quasiprobability distribution function in phase space, which itself is derived from the master equation for the quantum density operator \cite{Drummond1980,GardinerCrispin1985}. The derivation relies on the assumption that the positive-$P$ distribution function is sufficiently bounded so as to lead to vanishing boundary terms in the integration-by-parts step of the derivation of the Fokker-Planck equation. For vanishing boundary terms, the positive-$P$ approach and the resulting SDEs are equivalent to simulating the \emph{exact} quantum dynamics of the system. Whether this assumption is actually satisfied or not can in practice be checked \emph{a posteriori, i.e.,} after carrying out the numerical simulations of the SDEs. Vanishing boundary terms lead to stable evolution, with sampling errors that can be controlled and reduced by increasing the number of stochastic realizations required to calculate the stochastic averages. Non-vanishing boundary terms, on the other hand, lead to the well-known occurrence of `spiking' trajectories and a development of power-law tails in the quasiprobability distribution function \cite{Gilchrist1997}. This in turn leads to rapid and uncontrollable growth of sampling errors as the simulation progresses in time \cite{Deuar2006}, implying that the simulation results can no longer be used past the `spiking' time.

The complex stochastic fields $\Psi(\mathbf{r},t)$ and $\widetilde{\Psi}(\mathbf{r},t)$ are independent of each other {[}$\widetilde{\Psi}(\mathbf{r},t)\neq\Psi^{\ast}(\mathbf{r},t)${]} except in the mean, $\langle\widetilde{\Psi}(\mathbf{r},t)\rangle=\langle\Psi^{\ast}(\mathbf{\mathbf{r}},t)\rangle$, where the brackets $\langle\ldots\rangle$ refer to stochastic averages with respect to the positive-$P$ distribution function. In numerical realizations, this is represented by an ensemble average over a large number of stochastic realizations (trajectories). Observables described by quantum mechanical ensemble averages over normally-ordered products of the field operators have an exact correspondence with stochastic averages over products of the $c$-fields $\Psi(\mathbf{r},t)$ and $\widetilde{\Psi}(\mathbf{r},t)$, such that 
\begin{equation}
	\langle\lbrack\hat{\Psi}^{\dagger}(\mathbf{r},t)]^{m}[\hat{\Psi}(\mathbf{r}^{\prime},t)]^{n}\rangle=\langle\lbrack\widetilde{\Psi}(\mathbf{r},t)]^{m}[\Psi(\mathbf{r}^{\prime},t)]^{n}\rangle.\label{eq:correspondence}
\end{equation}

Converting the above SDEs from  \^{I}to to Stratonovich form, which has the advantage of obeying the standard rules of calculus, leads only to a physically irrelevant constant phase shift term which can be ignored. Thus, the Stratonovich form of the SDEs is the same as the above  \^{I}to form. Once the Stratonovich form of the SDEs is established, we can transform to a new pair of complex variables -- `density' and `phase' -- defined via 
\begin{align}
	\rho(\mathbf{r},t) & =\widetilde{\Psi}(\mathbf{r},t)\Psi(\mathbf{r},t),\\
	S(\mathbf{r},t) & =\frac{1}{2i}\ln\frac{\Psi(\mathbf{r},t)}{\widetilde{\Psi}(\mathbf{r},t)}=\frac{1}{2i}\left(\ln\Psi(\mathbf{r},t)-\ln\widetilde{\Psi}(\mathbf{r},t)\right).\label{eq:density-phase}
\end{align}
These transformations correspond to 
\begin{align}
	\Psi(\mathbf{r},t) & =\sqrt{\rho(\mathbf{r},t)}e^{iS(\mathbf{r},t)},\label{eq:fields-density-phase1}\\
	\widetilde{\Psi}(\mathbf{r},t) & =\sqrt{\rho(\mathbf{r},t)}e^{-iS(\mathbf{r},t)}.\label{eq:fields-density-phase2}
\end{align}

The new SDEs for the complex density and phase variables can be written as 
\begin{align}
	\frac{\partial\rho}{\partial t} & =-\frac{\hbar}{m}\boldsymbol{\nabla}\cdot(\rho\boldsymbol{\nabla}S)+\sqrt{-i\frac{g}{\hbar}}\rho\left(\xi_{1}+i\xi_{2}\right),\\
	\frac{\partial S}{\partial t} & =-\frac{\hbar}{2m}(\boldsymbol{\nabla}S)^{2}-\frac{V_{\mathrm{ext}}}{\hbar}-\frac{g}{\hbar}\rho+\frac{\hbar}{2m\sqrt{\rho}}\laplacian\sqrt{\rho}\nonumber \\
	& \quad\quad\quad\quad\quad\quad\quad\quad-\frac{1}{2}\sqrt{i\frac{g}{\hbar}}\left(\xi_{1}-i\xi_{2}\right),\label{eq:SQHD-c}
\end{align}
where the term involving $\laplacian\sqrt{\rho}$ can be alternatively rewritten as 
\begin{equation}
	\frac{1}{\sqrt{\rho}}\laplacian\sqrt{\rho}=\frac{1}{2\rho}\laplacian\rho-\frac{1}{4\rho^{2}}(\boldsymbol{\nabla}\rho)^{2}=\frac{1}{2}\laplacian\ln\rho+\frac{1}{4}(\boldsymbol{\nabla}\ln\rho)^{2}.
\end{equation}

Note that the noise term in the equation for the density field $\rho$ is \emph{multiplicative} (\emph{i.e}., it depends on the stochastic variable itself), whereas the noise term in the equation for the phase field $S$ is \emph{additive} and therefore it corresponds to a random stochastic process that has Gaussian statistics. The Gaussian statistics for the phase variable implies that, for example, stochastic averages of the form $\left\langle e^{iS}\right\rangle$ can be expressed as 
\begin{equation}
	\langle e^{iS}\rangle=e^{-\frac{1}{2}\langle S^{2}\rangle}.
\end{equation}
This step, however, is not required from a numerical point of view as the stochastic average value of $\langle e^{iS}\rangle$ can be evaluated explicitly just as easily as $\langle S^{2}\rangle$.

The final step in deriving the stochastic  hydrodynamic equations involves the introduction of a (complex) stochastic `velocity' vector field
\begin{equation}
	\boldsymbol{v}(\mathbf{r},t)=\frac{\hbar}{m}\boldsymbol{\nabla}S(\mathbf{r},t),\label{eq:velocity-def}
\end{equation}
which is analogous to the standard velocity vector in conventional hydrodynamics: after taking the stochastic average, the expectation value of the stochastic velocity vector becomes real and acquires the same physical meaning as in conventional hydrodynamics. This leads us to the following SQHD equations,
\begin{align}
	\frac{\partial\rho}{\partial t} & =-\boldsymbol{\nabla}\cdot(\rho\boldsymbol{v})+\sqrt{-i\frac{g}{\hbar}}\rho\left(\xi_{1}+i\xi_{2}\right),\label{eq:SQHD-a}\\
	\frac{\partial\boldsymbol{v}}{\partial t} & =-\boldsymbol{\nabla}\left[\frac{1}{2}\boldsymbol{v}^{2}+\frac{V_{\mathrm{ext}}}{m}+\frac{g}{m}\rho-\frac{\hbar^{2}}{2m^{2}\sqrt{\rho}}\laplacian\sqrt{\rho}\right.\nonumber \\
 	& \left.\quad\quad\quad\quad\quad\quad\quad\quad\quad+\frac{1}{2m}\sqrt{i\hbar g}\left(\xi_{1}-i\xi_{2}\right)\right],\label{eq:SQHD-b}
\end{align}
which are the first key result of this work. 

As we will show below, the utility of these equations is three-fold; (\emph{i}) they provide a theoretically exact description of interacting Bose gases in terms of hydrodynamic variables; (\emph{ii}) they can be used to derive new approximate hydrodynamic schemes, where connections to alternative theories can be directly shown; and (\emph{iii}) such hydrodynamic theories can be advantageous over alternatives in a number of different physical scenarios.

\section{Approximate schemes in stochastic quantum hydrodynamics}

\subsection{Suppressed density and phase fluctuations: hydrodynamics of pure condensates}

In this section we would like to apply the SQHD approach to Bose-Einstein condensates characterized by suppressed fluctuations of both the density and phase such that they can be ignored altogether. The goal is to derive the mean-field superfluid hydrodynamic equations directly from the stochastic hydrodynamic equations, providing an alternate derivation to the usual spontaneously broken symmetry approach (which leads to the Gross-Pitaevskii equation for the order parameter, and equations for the mean-field density and phase or velocity through the use of Madelung's transformation) \cite{pitaevskii_stringari_2016,Pethick&Smith2008,Castin2001}.

By taking the expectation values of both sides of Eqs. \eqref{eq:SQHD-a} and \eqref{eq:SQHD-b}, and taking into account that the noise terms $\xi_{i}$ are independent of the stochastic variables and have zero mean, we see that these equations will reduce to hydrodynamic equations for the mean density and the mean velocity, 
\begin{align}
	\frac{\partial\langle\rho\rangle}{\partial t} & =-\boldsymbol{\nabla}\cdot(\langle\rho\rangle\langle\boldsymbol{v}\rangle),\label{eq:SQHD-a_avg}\\
	\frac{\partial\langle\boldsymbol{v}\rangle}{\partial t} & =-\boldsymbol{\nabla}\left[\frac{1}{2}\langle\boldsymbol{v}\rangle^{2}+\frac{V_{\mathrm{ext}}}{m}+\frac{g}{m}\langle\rho\rangle\right.\nonumber\\
	&\qquad\qquad\qquad\left.-\frac{\hbar^{2}}{2m^{2}\sqrt{\langle\rho\rangle}}\laplacian\sqrt{\langle\rho\rangle}\right],\label{eq:SQHD-b_avg}
\end{align}
if we assume that:

(\emph{i}) the average of products of stochastic velocities (phases) factorize into products of average velocities (phases), so that $\langle\boldsymbol{v}^{2}\rangle=\langle\boldsymbol{v}\rangle^{2}$. Such a condition implies that there are no longer any local velocity fluctuations (and therefore any local phase fluctuations) in the system since $\sigma_{\boldsymbol{v}}=\sqrt{\langle\boldsymbol{v}^{2}\rangle-\langle\boldsymbol{v}\rangle^{2}}=0$;

(\emph{ii}) the stochastic density and  velocity (phase) variables are uncorrelated, leading to the replacement $\langle\rho\boldsymbol{v}\rangle=\langle\rho\rangle\langle\boldsymbol{v}\rangle$. This follows from the condition above, since there are no velocity (phase) fluctuations;

(\emph{iii}) the average of products or ratios of the stochastic densities factorize into respective products or ratios of average densities, so that the term $\langle\frac{1}{\sqrt{\rho}}\laplacian\sqrt{\rho}\rangle$ can be replaced by $\frac{1}{\sqrt{\langle\rho\rangle}}\laplacian\sqrt{\langle\rho\rangle}$.

Assumptions (\emph{i})--(\emph{iii}) imply that all higher order moments of local operators factorize, leaving us with a coherent state, or equivalently, a mean-field description. This is similar to Glauber's definition of a coherent state of an optical field as a factorization property of all higher-order correlation functions, equivalent to its definition as an eigenstate of the field annihilation operator  \cite{glauber2007quantum,Glauber_1963_II,Glauber_1963_I}.  %\cite{Glauber_Footnote}.
It is satisfying to note then, that this alternate derivation has led us back to equations of the same form as the usual mean-field superfluid hydrodynamic equations \cite{pitaevskii_stringari_2016,Pethick&Smith2008,Castin2001},
\begin{align}
	\frac{\partial\rho_{0}}{\partial t} & =-\boldsymbol{\nabla}\cdot(\rho_{0}\boldsymbol{v}_{0}),\label{eq:QHD-a}\\
	\frac{\partial\boldsymbol{v}_{0}}{\partial t} & =-\boldsymbol{\nabla}\left(\frac{1}{2}\boldsymbol{v}_{0}^{2}+\frac{V_{\mathrm{ext}}}{m}+\frac{g}{m}\rho_{0}-\frac{\hbar^{2}}{2m^{2}\sqrt{\rho_{0}}}\laplacian\sqrt{\rho_{0}}\right),\label{eq:QHD-b}\\
	\frac{\partial S_{0}}{\partial t} & =-\frac{\hbar}{2m}(\boldsymbol{\nabla}S_{0})^{2}-\frac{V_{\mathrm{ext}}}{\hbar}-\frac{g}{\hbar}\rho_{0}+\frac{\hbar}{2m\sqrt{\rho_{0}}}\laplacian\sqrt{\rho_{0}},\label{eq:QHD-c}
\end{align}
where one can identify that,
\begin{equation}
	\langle\rho\rangle=\rho_{0},\quad\langle\boldsymbol{v}\rangle=\boldsymbol{v}_{0},\quad\langle S\rangle=S_{0}.
\end{equation}

Equations \eqref{eq:QHD-a}-\eqref{eq:QHD-c} of superfluid hydrodynamics are usually derived in the spontaneously broken symmetry approach by applying Madelung's transformation $\Psi_{0}=\sqrt{\rho_{0}}e^{iS_{0}}$ to the order parameter  $\Psi_{0}=\langle\hat{\Psi}\rangle=\langle\Psi\rangle$, with $\Psi_{0}^{*}=\langle\hat{\Psi}^{\dagger}\rangle=\langle\tilde{\Psi}\rangle$, along with the definition $\boldsymbol{v}_{0}=\frac{\hbar}{m}\boldsymbol{\nabla}S_{0}$, where the order parameter itself evolves according to the mean-field Gross-Pitaevskii equation (GPE),
\begin{align}
	\frac{\partial\Psi_{0}}{\partial t}&=\frac{i}{\hbar}\left[\frac{\hbar^{2}}{2m}\laplacian-V_{\text{ext}}-g|\Psi_{0}|^{2}\right]\Psi_{0}.\label{eq:GPE}
\end{align}
%\Rev{We note that this usual derivation relies on the assumption of long-range order, whereas our new alternate derivation instead relies on the explicit factorization of density and phase/velocity correlations.}

When dealing with the superfluid hydrodynamic equations \eqref{eq:QHD-a}-\eqref{eq:QHD-c}, the quantum pressure term $\frac{1}{\sqrt{\rho_{0}}}\laplacian\sqrt{\rho_{0}}$ is often neglected in the literature since it is responsible for dynamics on short length scales $\sim l_{\text{h}}=\hbar/\sqrt{mg\rho_{0}}$ (the healing length) and can be ignored when considering the bulk dynamics of a system at large scales \cite{Cazalilla2011,Damski2004}. Removing this term results in the classical Euler hydrodynamic equations for the density and velocity,
\begin{align}
	\frac{\partial\rho_{0}}{\partial t} & =-\boldsymbol{\nabla}\cdot(\rho_{0}\boldsymbol{v}_{0}),\label{eq:CHD-a}\\
	\frac{\partial\boldsymbol{v}_{0}}{\partial t} & =-\boldsymbol{\nabla}\left(\frac{1}{2}\boldsymbol{v}_{0}^{2}+\frac{V_{\mathrm{ext}}}{m}\right) - \frac{1}{m \rho_{0}} \boldsymbol{\nabla} P.\label{eq:CHD-b}
\end{align}
Such equations can be postulated phenomenologically by considering the dynamics of small uniform slices of the Bose gas and employing the local density approximation. Here we have written the equations in standard form where $P=g\rho_{0}^{2}/2$ is the pressure (or equation of state) of a weakly interacting Bose gas at zero temperature \cite{pitaevskii_stringari_2016}, related to the chemical potential $\mu=g\rho_{0}$ via the Gibbs-Duhem relation $dP=\rho_{0}d\mu$ \cite{Pethick&Smith2008}.

\subsection{Linearized stochastic quantum hydrodynamics} \label{sec:Linearised stochastic quantum hydrodynamics}

The next typical situation we would like to address via the stochastic quantum hydrodynamic approach is the description of the dynamics of quantum fluids that have small density and phase fluctuations. We would like to simplify the stochastic equations \eqref{eq:SQHD-a} and \eqref{eq:SQHD-b} in such a way that the \emph{multiplicative} noise terms (which are known to be the source of uncontrollable sampling errors past the `spiking' time \cite{Gilchrist1997,Deuar2006}) can be approximated by \emph{additive} noise. This can be done in a linearized treatment of fluctuations around their deterministic (mean-field) value by assuming that the fluctuations are small and truncating the relevant expansion terms in the equations of motion. The assumption that density and phase fluctuations are small is valid for 3D weakly-interacting Bose-Einstein condensates where true long-range order exists. However, when considering 1D and 2D systems more care must be taken, since phase-fluctuations persist at low temperatures (even at zero temperature in 1D) and true long-range order is absent for uniform systems in the thermodynamic limit \cite{Hohenberg1967,petrov2004low,Cazalilla2004,Castin2004}. Nevertheless, for finite-size trapped systems it is still possible to employ a linearized treatment of fluctuations in these cases, provided that the first order correlation function decays slowly enough, or in other words, that the system size $L$ is smaller than the relevant phase-coherence length $l_{\phi}$ so that long-range order is recovered within the system \cite{petrov2004low,Petrov2000,Bouchoule2012,Morgan_2000,Damski2006,Mathey2009,Imambekov2009}.

Thus, we proceed by decomposing the density, velocity, and phase variables as a sum of two components: a real-valued deterministic component that obeys the superfluid hydrodynamic equations, and a complex-valued fluctuating component that describes any perturbations on top of the mean-field description. We denote such a decomposition using,
% the deterministic (mean-field) components and the fluctuating stochastic components,
% \begin{align}
% 	\rho & =\left\langle \rho\right\rangle +\delta\rho=\rho_{0}+\delta\rho,\\
% 	\boldsymbol{v} & =\left\langle \boldsymbol{v}\right\rangle +\delta\boldsymbol{v}=\boldsymbol{v}_{0}+\delta\boldsymbol{v},\\
% 	S & =\left\langle S\right\rangle +\delta S=S_{0}+\delta S.
% \end{align}
\begin{align}
	\rho(\mathbf{r},t) &=\rho_{0}(\mathbf{r},t)+\delta\rho(\mathbf{r},t),\label{eq:decomp_rho}\\
	\boldsymbol{v}(\mathbf{r},t) &=\boldsymbol{v}_{0}(\mathbf{r},t)+\delta\boldsymbol{v}(\mathbf{r},t),\label{eq:decomp_v}\\
	S(\mathbf{r},t) &=S_{0}(\mathbf{r},t)+\delta S(\mathbf{r},t)\label{eq:decomp_S},
\end{align}
where $\rho_{0}(\mathbf{r},t)$, $\boldsymbol{v}_{0}(\mathbf{r},t)$, $S_{0}(\mathbf{r},t)$ obey Eqs. \eqref{eq:QHD-a}--\eqref{eq:QHD-c}, and the evolution equations for $\delta\rho(\mathbf{r},t)$, $\delta\boldsymbol{v}(\mathbf{r},t)$, $\delta S(\mathbf{r},t)$ are yet to be determined.

Taking the expectation value of the SQHD equations \eqref{eq:SQHD-a}, \eqref{eq:SQHD-b} and \eqref{eq:SQHD-c}, one can show that, to linear order in fluctuating quantities, the superfluid hydrodynamic equations exactly capture the mean behavior of such a description, \textit{i.e.} $\langle \rho(\mathbf{r},t)\rangle=\rho_{0}(\mathbf{r},t)$, $\langle \boldsymbol{v}(\mathbf{r},t)\rangle=\boldsymbol{v}_{0}(\mathbf{r},t)$ and $\langle S(\mathbf{r},t)\rangle=S_{0}(\mathbf{r},t)$. This means that, at linear order, $\langle\delta\rho(\mathbf{r},t)\rangle=\langle\delta\boldsymbol{v}(\mathbf{r},t)\rangle=\langle\delta S(\mathbf{r},t)\rangle=0$. Therefore, while a linearized scheme cannot provide corrections to quantities like the real-space density, it instead finds its primary utility in the ability to calculate quantum correlation functions in dynamical scenarios, since leading order corrections can be kept when considering these observables (see Sec. \ref{sec:In the linearised treatment of quantum fluctuations}). We derive such linearized equations shortly, and calculate correlation functions later in this paper, however we pause briefly now to consider fluctuations up to second order. In this case $\langle\delta\rho(\mathbf{r},t)\rangle,\langle\delta\boldsymbol{v}(\mathbf{r},t)\rangle,\langle\delta S(\mathbf{r},t)\rangle\neq0$ and it is therefore possible to obtain quantum corrections to the mean-field density, velocity and phase. Such a second-order scheme remains valid provided that perturbations from the mean-field result can be considered small, that is, $\langle\delta\rho(\mathbf{r},t)\rangle\ll\rho_{0}(\mathbf{r},t)$, $\langle\delta\boldsymbol{v}(\mathbf{r},t)\rangle\ll\boldsymbol{v}_{0}(\mathbf{r},t)$, $\langle\delta S(\mathbf{r},t)\rangle\ll S_{0}(\mathbf{r},t)$. The equations governing the fluctuating fields can be derived by substituting the decompositions \eqref{eq:decomp_rho}--\eqref{eq:decomp_S} into the SQHD equations [\eqref{eq:SQHD-a}, \eqref{eq:SQHD-b} and \eqref{eq:SQHD-c}] and employing the superfluid hydrodynamic equations \eqref{eq:QHD-a}--\eqref{eq:QHD-c} for the time evolution of the deterministic components. Then, truncating to second order in fluctuating quantities leads to the following SDEs:
% The deterministic components are described by the superfluid hydrodynamic equations, Eqs. \eqref{eq:QHD-a}--\eqref{eq:QHD-c}, whereas the complex fluctuating components are described to second order by the following SDEs: 
\begin{align}
	\frac{\partial\delta\rho}{\partial t} & =-\boldsymbol{\nabla}\cdot\left[\rho_{0}\delta\boldsymbol{v}+(\boldsymbol{v}_{0}+\delta\boldsymbol{v})\delta\rho\right]\nonumber \\
	& \qquad\qquad\qquad+\sqrt{-i\frac{g}{\hbar}}(\rho_{0}+\delta\rho)\,(\xi_{1}+i\xi_{2}),\label{eq:L2SQH-a}
\end{align}
\begin{align}
	\frac{\partial\delta\boldsymbol{v}}{\partial t} & =-\boldsymbol{\nabla}\left[(\boldsymbol{v}_{0}\cdot\delta\boldsymbol{v})+\frac{1}{2}\delta\boldsymbol{v}^{2}+\frac{g}{m}\delta\rho\right.\nonumber \\
	& \quad-\frac{\hbar^{2}}{2m^{2}}\left\{-\frac{1}{2}\frac{\delta\rho}{\rho_{0}^{3/2}}\laplacian\sqrt{\rho_{0}}+\frac{1}{2}\frac{1}{\sqrt{\rho_{0}}}\laplacian\left(\frac{\delta\rho}{\sqrt{\rho_{0}}}\right)\right.\nonumber\\
	&\quad+\frac{3}{8}\left(\frac{\delta \rho}{\rho_{0}^{5/4}}\right)^{2}\laplacian\sqrt{\rho_{0}}-\frac{1}{4}\frac{\delta\rho}{\rho_{0}^{3/2}}\laplacian\left(\frac{\delta\rho}{\sqrt{\rho_{0}}}\right)\nonumber\\
	&\quad\left.\left.-\frac{1}{8}\frac{1}{\sqrt{\rho_{0}}}\laplacian\left(\frac{\delta\rho}{\rho_{0}^{3/4}}\right)^{2}+\mathcal{O}(\delta\rho^{3})\right\}\right.\nonumber\\
	&\left.\qquad\qquad\qquad\qquad\quad+\frac{1}{2m}\sqrt{i\hbar g}\,(\xi_{1}-i\xi_{2})\right],\label{eq:L2SQH-b}
\end{align}
\begin{align}
	\frac{\partial\mathcal{\delta S}}{\partial t} & =-\frac{\hbar}{m}(\boldsymbol{\nabla}S_{0}\cdot\boldsymbol{\nabla}\delta S)-\frac{\hbar}{2m}(\boldsymbol{\nabla}\delta S)^{2}-\frac{g}{\hbar}\delta\rho\nonumber\\
	& \quad+\frac{\hbar}{2m}\left\{-\frac{1}{2}\frac{\delta\rho}{\rho_{0}^{3/2}}\laplacian\sqrt{\rho_{0}}+\frac{1}{2}\frac{1}{\sqrt{\rho_{0}}}\laplacian\left(\frac{\delta\rho}{\sqrt{\rho_{0}}}\right)\right.\nonumber\\
	&\quad+\frac{3}{8}\left(\frac{\delta \rho}{\rho_{0}^{5/4}}\right)^{2}\laplacian\sqrt{\rho_{0}}-\frac{1}{4}\frac{\delta\rho}{\rho_{0}^{3/2}}\laplacian\left(\frac{\delta\rho}{\sqrt{\rho_{0}}}\right)\nonumber\\
	&\quad-\left.\frac{1}{8}\frac{1}{\sqrt{\rho_{0}}}\laplacian\left(\frac{\delta\rho}{\rho_{0}^{3/4}}\right)^{2}+\mathcal{O}(\delta\rho^{3})\right\}\nonumber\\
	&\qquad\qquad\qquad\qquad\quad-\frac{1}{2}\sqrt{i\frac{g}{\hbar}}\,(\xi_{1}-i\xi_{2}), \label{eq:L2SQH-c}
\end{align}
where the relationship between $\delta S$ and $\delta\boldsymbol{v}$ is given by $\delta\boldsymbol{v}=\frac{\hbar}{m}\boldsymbol{\nabla}\delta S$, similarly to the usual $\boldsymbol{v}_{0}=\frac{\hbar}{m}\boldsymbol{\nabla}S_{0}$. We keep the phase equation here since it can be a more convenient variable to simulate for the computation of certain observables (see Sec. \ref{sec: In the full stochastic quantum hydrodynamics}), whereas the velocity equation provides the more natural set of SDEs from a hydrodynamic perspective.

The form of these equations above is particularly helpful in highlighting the role of the stochastic many-body quantum pressure term $\frac{1}{\sqrt{\rho}}\laplacian\sqrt{\rho}$ in the full SQHD equations. While formally similar to the mean-field quantum pressure $\frac{1}{\sqrt{\rho_{0}}}\laplacian\sqrt{\rho_{0}}$, the many-body quantum pressure term contains much more information, incorporating high order density fluctuations relevant to the many-body nature of interacting Bose gases. This additional information contained within the many-body quantum pressure term can be seen in equations \eqref{eq:L2SQH-b} and \eqref{eq:L2SQH-c} where the terms inside $\left\{\cdots\right\}$ describe the remaining higher order density fluctuations after truncating the infinite series expansion down to second order. While in this work we are going to further restrict ourselves to first order in fluctuations, in principle one can incorporate these second and even higher order expansion terms in the many-body quantum pressure as an avenue for future developments (see Sec. \ref{Conclusions} below). 

We also note here that, at second order in the fluctuating components, the many-body quantum pressure is the only term that has been truncated. All other terms do not contribute anything higher than second order in fluctuations, \textit{i.e.} no terms involving the phase or velocity field have been neglected in equations \eqref{eq:L2SQH-a}-\eqref{eq:L2SQH-c}. As a foreshadow of Section \ref{sec:Comparison with Bogoliubov approaches} below, we mention that keeping these second order terms corresponds to a description which goes beyond the Bogoliubov treatment of interacting Bose gases. Again, we leave this open as an interesting direction to pursue in future work.

Proceeding further and truncating down to linear order, we arrive at
\begin{align}
	\frac{\partial\delta\rho}{\partial t} & =-\boldsymbol{\nabla}\cdot\left[\rho_{0}\delta\boldsymbol{v}+\boldsymbol{v}_{0}\delta\rho\right]+\sqrt{-i\frac{g}{\hbar}}\rho_{0}\,(\xi_{1}+i\xi_{2}),\label{eq:LSQHD-a}\\
	\frac{\partial\delta\boldsymbol{v}}{\partial t} & =-\boldsymbol{\nabla}\left[(\boldsymbol{v}_{0}\cdot\delta\boldsymbol{v})+\frac{g}{m}\delta\rho-\frac{\hbar^{2}}{4m^{2}\sqrt{\rho_{0}}}\laplacian\left(\frac{\delta\rho}{\sqrt{\rho_{0}}}\right)\right.\nonumber \\
	& \quad+\left.\frac{\hbar^{2}}{4m^{2}}\frac{\delta\rho}{\rho_{0}^{3/2}}
	\laplacian\sqrt{\rho_{0}}+\frac{1}{2m}\sqrt{i\hbar g}\,(\xi_{1}-i\xi_{2})\right],\label{eq:LSQHD-b}\\
	\frac{\partial\mathcal{\delta S}}{\partial t} & =-\frac{\hbar}{m}(\boldsymbol{\nabla}S_{0}\cdot\boldsymbol{\nabla}\delta S)-\frac{g}{\hbar}\delta\rho+\frac{\hbar}{4m\sqrt{\rho_{0}}}\laplacian\left(\frac{\delta\rho}{\sqrt{\rho_{0}}}\right)\nonumber \\
	& \quad-\frac{\hbar}{4m}\frac{\delta\rho}{\rho_{0}^{3/2}}\laplacian\sqrt{\rho_{0}}-\frac{1}{2}\sqrt{i\frac{g}{\hbar}}\,(\xi_{1}-i\xi_{2}).\label{eq:LSQHD-c}
\end{align}

As we see, all fluctuating noise terms are now additive as the contribution of terms involving products of stochastic components has been neglected on the grounds of being higher-than-linear order in small fluctuations. 
By small here we mean that $\langle\delta\rho^{2}(\mathbf{r},t)\rangle\ll\rho_{0}^{2}(\mathbf{r},t)$, $\langle\delta\boldsymbol{v}^{2}(\mathbf{r},t)\rangle\ll\boldsymbol{v}_{0}^{2}(\mathbf{r},t)$, and $\langle\delta S^{2}(\mathbf{r},t)\rangle\ll S_{0}^{2}(\mathbf{r},t)$, since $\langle\delta\rho(\mathbf{r},t)\rangle=\langle\delta\boldsymbol{v}(\mathbf{r},t)\rangle=\langle\delta S(\mathbf{r},t)\rangle=0$ in this case (as mentioned previously).
% By small here we mean that $\langle\delta\rho(\mathbf{r},t)\rangle\ll\rho_{0}(\mathbf{r},t)$, $\langle\delta\boldsymbol{v}(\mathbf{r},t)\rangle\ll\boldsymbol{v}_{0}(\mathbf{r},t)$, and $\langle\delta S(\mathbf{r},t)\rangle\ll S_{0}(\mathbf{r},t)$. In the case when $\langle\delta\rho(\mathbf{r},t)\rangle=0$, $\langle\delta\boldsymbol{v}(\mathbf{r},t)\rangle=0$, $\langle\delta S(\mathbf{r},t)\rangle=0$, then one would consider $\langle\delta\rho^{2}(\mathbf{r},t)\rangle\ll\rho_{0}^{2}(\mathbf{r},t)$, $\langle\delta\boldsymbol{v}^{2}(\mathbf{r},t)\rangle\ll\boldsymbol{v}_{0}^{2}(\mathbf{r},t)$, and $\langle\delta S^{2}(\mathbf{r},t)\rangle\ll S_{0}^{2}(\mathbf{r},t)$.

The above LSQHD equations \eqref{eq:LSQHD-a}--\eqref{eq:LSQHD-c}, which have to be solved numerically in conjunction with the mean-field equations \eqref{eq:QHD-a}-\eqref{eq:QHD-c}, are the second key result of this work. Although they are approximate, their advantage over the full SQHD equations lies in the fact that they remain stable and can be simulated for significantly longer times (and remain valid provided the fluctuations do not grow too large) due to the \emph{additive}  nature of the noise terms, as opposed to the \emph{multiplicative} noise appearing in the full SQHD equations.

\subsection{Comparison with Bogoliubov approaches} \label{sec:Comparison with Bogoliubov approaches}
To elucidate the utility of the LSQHD equations \eqref{eq:LSQHD-a}--\eqref{eq:LSQHD-c}, and before turning to numerical examples, we now explore their comparison with other known numerical methods in the literature. In particular, since these equations are derived under a linearized scheme of quantum fluctuations, we focus on connections to Bogoliubov type theories.

Specifically, under the transformations
	\begin{align}
		\delta\psi&=\Psi_{0}\left(\frac{\delta\rho}{2\rho_{0}}+i\delta S\right)=\sqrt{\rho_{0}}e^{iS_{0}}\left(\frac{\delta\rho}{2\rho_{0}}+i\delta S\right),\label{eq:DeltaPsi_HydroTransform}\\
		\delta\tilde{\psi}&=\Psi_{0}^{*}\left(\frac{\delta\rho}{2\rho_{0}}-i\delta S\right)=\sqrt{\rho_{0}}e^{-iS_{0}}\left(\frac{\delta\rho}{2\rho_{0}}-i\delta S\right),\label{eq:DeltaPsiTilde_HydroTransform}
	\end{align}
which correspond to $\delta\rho=\Psi_{0}^{*}\delta\psi+\Psi_{0}\delta\tilde{\psi}$ and $\delta S=\frac{1}{2i}\left[\frac{\delta\psi}{\Psi_{0}}-\frac{\delta\tilde{\psi}}{\Psi_{0}^{*}}\right]$, the LSQHD equations become
\begin{align}
	\frac{\partial\delta\psi(\mathbf{r},t)}{\partial t}&=\frac{i}{\hbar}\left[\frac{\hbar^{2}}{2m}\laplacian-V_{\text{ext}}(\mathbf{r},t)-2g|\Psi_{0}|^{2}\right]\delta\psi(\mathbf{r},t)\nonumber\\
	&\quad-\frac{i}{\hbar}g\Psi_{0}^{2}\delta\tilde{\psi}(\mathbf{r},t)+\sqrt{-i\frac{g}{\hbar}}\Psi_{0}\,\xi_{1}(\mathbf{r},t),\label{eq:SB-a}\\
	\frac{\partial\delta\tilde{\psi}(\mathbf{r},t)}{\partial t}&=-\frac{i}{\hbar}\left[\frac{\hbar^{2}}{2m}\laplacian-V_{\text{ext}}(\mathbf{r},t)-2g|\Psi_{0}|^{2}\right]\delta\tilde{\psi}(\mathbf{r},t)\nonumber\\
	&\quad+\frac{i}{\hbar}g(\Psi_{0}^{*})^{2}\delta\psi(\mathbf{r},t)+\sqrt{i\frac{g}{\hbar}}\Psi_{0}^{*}\,\xi_{2}(\mathbf{r},t),\label{eq:SB-b}
\end{align}
where $\Psi_{0}(\mathbf{r},t)$ is the mean-field order parameter that evolves according to the GPE \eqref{eq:GPE}.

Equations \eqref{eq:SB-a} and \eqref{eq:SB-b} are equivalent to the stochastic Bogoliubov equations first introduced in Ref. \cite{Kheruntsyan2010} (see also \cite{Ito_vs_Stratonovich}). Such equations have been used to successfully describe several experimental results on Bose-Einstein condensate collisions \cite{Kheruntsyan2010,Jaskula2010,Deuar2011,Kheruntsyan2012,Deuar2013,Kheruntsyan2014,Lewis-Swan-HOM}, and a more recent extension of this approach was used to characterize the quantum depletion of an expanding condensate \cite{Ross2022}. We point out that the stochastic Bogoliubov equations themselves are equivalent to the usual Bogoliubov approach \cite{pitaevskii_stringari_2016,Pethick&Smith2008,Bogoliubov1947,Pines1990} in which the Bose field operator is expanded as $\hat{\Psi}=\Psi_{0}+\delta\hat{\psi}$ where $\Psi_{0}$ is the mean-field order parameter obeying equation \eqref{eq:GPE} and $\delta\hat{\psi}$ represents small fluctuations around the mean-field obeying
\begin{align}
	\frac{\partial\delta\hat{\psi}(\mathbf{r},t)}{\partial t}&=\frac{i}{\hbar}\left[\frac{\hbar^{2}}{2m}\laplacian-V_{\text{ext}}(\mathbf{r},t)-2g|\Psi_{0}|^{2}\right]\delta\hat{\psi}(\mathbf{r},t)\nonumber\\
	&\qquad\qquad\qquad\quad-\frac{i}{\hbar}g\Psi_{0}^{2}\delta\hat{\psi}^{\dagger}(\mathbf{r},t).\label{eq:Bogo-delta}
\end{align}
Hence, we find that the evolution equations of LSQHD are equivalent to the stochastic Bogoliubov equations as well as the usual Bogoliubov approach, which proceeds through the diagonalization of $\delta\hat{\psi}$ via the mode decomposition $\delta\hat{\psi}(\mathbf{r},t)=e^{-i\mu t/\hbar}\sum_{i=1}^{\infty}[u_{i}(\mathbf{r},t)\hat{a}_{i}+v_{i}^{*}(\mathbf{r},t)\hat{a}_{i}^{\dagger}]$ leading to the well-known time-dependent Bogoliubov--de Gennes (BdG) equations \cite{Fetter1972,deGennes1999,Dziarmaga2005}, where $\mu$ is the chemical potential of the condensate mode, $u_{i}$ and $v_{i}$ are the Bogoliubov mode amplitudes, and $\hat{a}_{i}^{\dagger}$ and $\hat{a}_{i}$ the respective quasiparticle creation and annihilation operators.

While the evolution equations of the LSQHD, stochastic Bogoliubov, and usual Bogoliubov approaches are equivalent, the initial state of each requires consideration. After diagonalization, the usual Bogoliubov approach begins in an initial Bogoliubov state which evolves according to the usual time-dependant BdG equations. By contrast, the stochastic schemes are usually initialized in a coherent state, since a more accurate description of the initial ground state of a Bose-Einstein condensate does not currently exist in the literature for zero temperature positive-\textit{P} approaches. This coherent state initial condition is achieved by simply not seeding the initial state with any noise. For the examples considered in this work, one begins with a desired mean-field initial state described by $\rho_{0}(\mathbf{r},0)$ and $\boldsymbol{v}_{0}(\mathbf{r},0)$ or $S_{0}(\mathbf{r},0)$ (equivalently $\Psi_{0}(\mathbf{r},0)$), and sets all initial fluctuating fields to zero: $\delta\rho(\mathbf{r},0)=\delta\boldsymbol{v}(\mathbf{r},0)=\delta S(\mathbf{r},0)=0$ or equally $\delta\psi(\mathbf{r},0)=\delta\tilde{\psi}(\mathbf{r},0)=0$. This is equivalent to initializing the full positive-\textit{P} equations with $\rho(\mathbf{r},0)=\rho_{0}(\mathbf{r},0)=|\Psi_{0}(\mathbf{r},0)|^{2}$, $\boldsymbol{v}(\mathbf{r},0)=\boldsymbol{v}_{0}(\mathbf{r},0)$, $S(\mathbf{r},0)=S_{0}(\mathbf{r},0)$ or $\Psi(\mathbf{r},0)=\Psi_{0}(\mathbf{r},0)$, $\widetilde{\Psi}(\mathbf{r},0)=\Psi_{0}^{*}(\mathbf{r},0)$. Such an initialization avoids the need for diagonalization entirely. Hence, for scenarios where exact diagonalization becomes computationally intractable, the stochastic methods become invaluable. Some situations where this type of advantage has already been exploited can be found in Refs. \cite{Kheruntsyan2010,Kheruntsyan2014,Ross2022}. In these cases, the systems being considered require an extremely large numerical lattice or an unreasonable number of modes for simulation. Hence, exact diagonalization in these types of situations is difficult and undesirable. An equivalent stochastic approach, like the one derived in this work, provides a means of avoiding this. While we restrict ourselves to zero temperature here, we note that it is also possible to consider finite temperature situations by stochastically sampling the appropriate thermal \textit{P}-distribution in order to construct the initial state \cite{Olsen2009,Lewis-Swan-qBEC}.

Additionally, we mention that since the LSQHD equations are Bogoliubov in nature (\textit{i.e.} result from a quadratic Hamiltonian), this means that all higher order correlations factorize into products of second order moments, according to Wick's theorem. At the level of field operators, this implies that we know everything about the system after computing only the normal and anomalous correlators, $G^{(1)}_{n}(\mathbf{r},\mathbf{r}^{\prime},t)\equiv\langle\delta\hat{\psi}^{\dagger}(\mathbf{r},t)\delta\hat{\psi}(\mathbf{r}^{\prime},t)\rangle=\langle\delta\tilde{\psi}(\mathbf{r},t) \delta\psi(\mathbf{r}^{\prime},t)\rangle$ and $G^{(1)}_{a}(\mathbf{r},\mathbf{r}^{\prime},t)\equiv\langle\delta\hat{\psi}(\mathbf{r},t)\delta\hat{\psi}(\mathbf{r}^{\prime},t)\rangle=\langle\delta\psi(\mathbf{r},t) \delta\psi(\mathbf{r}^{\prime},t)\rangle$, respectively. In the LSQHD formalism this is equivalent to having knowledge of the correlators $\langle\delta\rho(\mathbf{r},t)\delta\rho(\mathbf{r}^{\prime},t)\rangle$, $\langle\delta S(\mathbf{r},t)\delta S(\mathbf{r}^{\prime},t)\rangle$, and  $\langle\delta\rho(\mathbf{r},t)\delta S(\mathbf{r}^{\prime},t)\rangle$. If one uses the transformations \eqref{eq:DeltaPsi_HydroTransform} and \eqref{eq:DeltaPsiTilde_HydroTransform} to convert the normal and anomalous correlators to hydrodynamic variables then this results in the relations
\begin{align}
	\langle&\delta\tilde{\psi}(\mathbf{r}) \delta\psi(\mathbf{r}^{\prime})\rangle\nonumber\\
	&= \sqrt{\rho_{0}(\mathbf{r})\rho_{0}(\mathbf{r}^{\prime})} e^{-i(S_{0}(\mathbf{r})-S_{0}(\mathbf{r}^{\prime}))} \nonumber\\
	&\hspace{5mm} \times \left[\frac{\langle\delta \rho(\mathbf{r}) \delta \rho(\mathbf{r}^{\prime})\rangle}{4 \rho_{0}(\mathbf{r}) \rho_{0}(\mathbf{r}^{\prime})} + \langle \delta S(\mathbf{r}) \delta S(\mathbf{r}^{\prime})\rangle \right.\nonumber\\
	&\hspace{11mm}+\left.\frac{i}{2} \left( \frac{\langle\delta \rho(\mathbf{r}) \delta S(\mathbf{r}^{\prime})\rangle}{\rho_{0}(\mathbf{r})} - \frac{\langle\delta S(\mathbf{r})\delta \rho(\mathbf{r}^{\prime})\rangle}{\rho_{0}(\mathbf{r}^{\prime})} \right)\right] \label{eq:g1_normal}
\end{align}
and
\begin{align}
	\langle&\delta\psi(\mathbf{r}) \delta\psi(\mathbf{r}^{\prime})\rangle\nonumber\\
	&= \sqrt{\rho_{0}(\mathbf{r})\rho_{0}(\mathbf{r}^{\prime})} e^{i(S_{0}(\mathbf{r})+S_{0}(\mathbf{r}^{\prime}))} \nonumber\\
	&\hspace{5mm}\times\left[\frac{\langle\delta \rho(\mathbf{r}) \delta \rho(\mathbf{r}^{\prime})\rangle}{4 \rho_{0}(\mathbf{r}) \rho_{0}(\mathbf{r}^{\prime})} - \langle \delta S(\mathbf{r}) \delta S(\mathbf{r}^{\prime})\rangle \right.\nonumber\\
	&\hspace{11mm}+\left.\frac{i}{2} \left( \frac{\langle\delta \rho(\mathbf{r}) \delta S(\mathbf{r}^{\prime})\rangle}{\rho_{0}(\mathbf{r})} + \frac{\langle\delta S(\mathbf{r})\delta \rho(\mathbf{r}^{\prime})\rangle}{\rho_{0}(\mathbf{r}^{\prime})}\right) \right] \label{eq:g1_anomalous}
\end{align}
where the time index has been omitted for brevity.

The diagonals of these correlators are then given by
\begin{align}
	\langle \delta\tilde{\psi}(\mathbf{r}) &\delta\psi(\mathbf{r})\rangle = \rho_{0}(\mathbf{r}) \left[\frac{\langle\delta\rho^{2}(\mathbf{r})\rangle}{4\rho_{0}^{2}(\mathbf{r})} + \langle\delta S^{2}(\mathbf{r})\rangle\right], \label{eq:diag_psi_tilde_psi}\\
	\langle \delta\psi(\mathbf{r})& \delta\psi(\mathbf{r})\rangle = \rho_{0}(\mathbf{r}) e^{2iS_{0}(\mathbf{r})}\nonumber\\
	&\times\left[\frac{\langle\delta\rho^{2}(\mathbf{r})\rangle}{4\rho_{0}^{2}(\mathbf{r})} + i \frac{\langle\delta\rho(\mathbf{r})\delta S(\mathbf{r})\rangle}{\rho_{0}(\mathbf{r})} - \langle\delta S^{2}(\mathbf{r})\rangle\right].
\end{align}

We take the opportunity now to point out a difference that arises between the Bogoliubov schemes and the LSQHD approach which we explore further in Sec. \ref{sec:In the linearised treatment of quantum fluctuations}. From an entirely hydrodynamic perspective, one would compute the real-space density within the LSQHD scheme using the diagonal of the reduced one-body density matrix $G^{(1)}(\mathbf{r},\mathbf{r},t) \equiv\langle\hat{\Psi}^{\dagger}(\mathbf{r},t)\hat{\Psi}(\mathbf{r},t)\rangle$ as
\begin{align}
	G^{(1)}(\mathbf{r},\mathbf{r},t) =\langle\widetilde{\Psi}(\mathbf{r},t)\Psi(\mathbf{r},t)\rangle
    =\left\langle\rho(\mathbf{r},t)\right\rangle 
	=\rho_{0}(\mathbf{r},t), \label{eq:G1rr_hydro}
\end{align}
since the superfluid hydrodynamic density $\rho_{0}(\mathbf{r},t)$ captures exactly the mean of $\rho(\mathbf{r},t)$ such that $\langle\delta\rho(\mathbf{r},t)\rangle=0$. Hence we see that the LSQHD approach can be considered a number conserving scheme which reproduces only the mean-field result for the real-space density. While the scheme is still capable of computing non-trivial results for higher order correlation functions, we recall that one must consider beyond-linear terms in the hydrodynamic equations of motion to obtain any corrections to the mean-field density itself.
% Due to the additive nature of the LSQHD equations, if they are initialized in a coherent state where $\delta\rho(\mathbf{r},0)=0$ and $\delta\boldsymbol{v}(\mathbf{r},0)=0$, like we do in this work, then the average density fluctuations will remain $\left\langle\delta\rho(\mathbf{r},t)\right\rangle=0$. Hence we see that the LSQHD approach can be considered a number conserving scheme which reproduces only the mean-field result for the real-space density. Whilst the scheme is still capable of computing non-trivial results for higher order correlation functions, one must consider beyond-linear terms in the hydrodynamic equations of motion to obtain any corrections to the mean-field density itself.

By contrast however, the usual Bogoliubov and stochastic Bogoliubov schemes presented here do provide corrections to the mean-field density in the sense that they are not number conserving. While more elaborate number conserving approaches are possible \cite{Gardiner1997,Castin1998,Dziarmaga2005,Billam2012,Billam2013,Ross2022}, the usual BdG equations and stochastic Bogoliubov scheme above rely on the undepleted pump approximation, where the number of particles in the mean-field or condensate mode $\Psi_{0}$ remains constant, and the number of particles in the excitations can vary. This can lead to dynamical population growth in the total particle number density of the system where the initial state of the fluctuating component is assumed to be a vacuum state. This is similar to spontaneous parametric down-conversion in quantum optics in the undepleted pump approximation, used as the simplest model for generation of the paradigmatic squeezed vacuum state \cite{walls2007quantum} (see also \cite{Ogren2009,KK_Fermionic}, and references therein).

Such population growth, along with the discrepancy between the Bogoliubov and hydrodynamic approaches, can be seen by examining the reduced one-body density matrix from the Bogoliubov perspective. In this case one expands the stochastic field using $\Psi=\Psi_{0}+\delta\psi$ (or $\widetilde{\Psi}=\Psi_{0}^{*}+\delta\tilde{\psi}$) and arrives at
\begin{align}
	G^{(1)}(\mathbf{r},&\mathbf{r},t) =\langle\widetilde{\Psi}(\mathbf{r},t)\Psi(\mathbf{r},t)\rangle\nonumber\\
	&\quad\,\,\,\,=|\Psi_{0}(\mathbf{r},t)|^{2}+\langle\delta\tilde{\psi}(\mathbf{r},t) \delta\psi(\mathbf{r},t)\rangle. \label{eq:G1rr_bogo}
\end{align}

% \begin{align}
% 	G^{(1)}(\mathbf{r},&\mathbf{r},t) =\langle\widetilde{\Psi}(\mathbf{r},t)\Psi(\mathbf{r},t)\rangle\nonumber\\
% 	&\quad\,\,\,\,=|\Psi_{0}(\mathbf{r},t)|^{2}+\Psi_{0}^{*}(\mathbf{r},t)\langle\delta\psi(\mathbf{r},t)\rangle\nonumber\\
% 	&+\Psi_{0}(\mathbf{r},t)\langle \delta\tilde{\psi}(\mathbf{r},t)\rangle+\langle\delta\tilde{\psi}(\mathbf{r},t) \delta\psi(\mathbf{r},t)\rangle. \label{eq:G1rr_bogo}
% \end{align}
Similarly to the LSQHD equations, since the stochastic Bogoliubov equations possess additive noise, the mean-field GPE describes exactly the average of the full stochastic field $\Psi(\mathbf{r},t)$ such that $\langle\Psi(\mathbf{r},t)\rangle=\Psi_{0}(\mathbf{r},t)$ and $\langle \delta\tilde{\psi}(\mathbf{r},t)\rangle=\langle \delta\psi(\mathbf{r},t)\rangle=0$. Equation \eqref{eq:G1rr_bogo} directly contrasts the hydrodynamic result $G^{(1)}(\mathbf{r},\mathbf{r},t)=\rho_{0}(\mathbf{r},t)=|\Psi_{0}(\mathbf{r},t)|^{2}$ of Eq. \eqref{eq:G1rr_hydro} for which the beyond-mean-field term $\langle\delta\tilde{\psi}(\mathbf{r},t) \delta\psi(\mathbf{r},t)\rangle$ (that is responsible for particle number growth) is missing. It is clear then, that inconsistencies will arise if one wants to convert Eq.~\eqref{eq:G1rr_bogo} into hydrodynamic variables using equation \eqref{eq:diag_psi_tilde_psi} where averages over second order density and phase variables appear. Such an inconsistency arises due to the fact that second order terms are often kept when dealing with observables like the real-space density, in order to obtain leading order corrections to the mean-field result. Yet the evolution equations themselves \eqref{eq:SB-a}, \eqref{eq:SB-b} and \eqref{eq:LSQHD-a}, \eqref{eq:LSQHD-b} along with the transformations \eqref{eq:DeltaPsi_HydroTransform}, \eqref{eq:DeltaPsiTilde_HydroTransform} are truncated only to first order.

% Similarly to the LSQHD equations, since the stochastic Bogoliubov equations are additive and we initialize them in a coherent state with $ \delta\tilde{\psi}(\mathbf{r},t)= \delta\psi(\mathbf{r},t)=0$, then the average of these fields remain $\langle \delta\tilde{\psi}(\mathbf{r},t)\rangle=\langle \delta\psi(\mathbf{r},t)\rangle=0$. This means one is left with $G^{(1)}(\mathbf{r},\mathbf{r},t)=|\Psi_{0}(\mathbf{r},t)|^{2}+\langle\delta\tilde{\psi}(\mathbf{r},t) \delta\psi(\mathbf{r},t)\rangle$ for the real-space density, which contrasts $G^{(1)}(\mathbf{r},\mathbf{r},t)=\rho_{0}(\mathbf{r},t)$ from Eq. \eqref{eq:G1rr_hydro}. The beyond-mean-field term here, $\langle\delta\tilde{\psi}(\mathbf{r},t) \delta\psi(\mathbf{r},t)\rangle$, is responsible for particle number growth in these approaches. It is clear then, that inconsistencies will arise if one wants to convert Eq.~\eqref{eq:G1rr_bogo} into hydrodynamic variables using equation \eqref{eq:diag_psi_tilde_psi} where averages over second order density and phase variables appear. Such an inconsistency arises due to the fact that second order terms are often kept when dealing with observables like the real-space density, in order to obtain leading order corrections to the mean-field result. Yet the evolution equations themselves \eqref{eq:SB-a}, \eqref{eq:SB-b} and \eqref{eq:LSQHD-a}, \eqref{eq:LSQHD-b} along with the transformations \eqref{eq:DeltaPsi_HydroTransform}, \eqref{eq:DeltaPsiTilde_HydroTransform} are truncated only to first order.

\subsection{Low-energy excitations and stochastic Luttinger liquid}

Finally, the last situation we consider in this section is that of low-energy excitations on top of an equilibrium density, which will lead to a Luttinger liquid description of a weakly-interacting Bose gas in 1D. To treat such a situation we begin with the full SQHD equations and write the stochastic density variable as $\rho=\rho_{\text{eq}}+\delta\rho$, where $\rho_{\text{eq}}$ represents the static equilibrium density and $\delta\rho$ any departure or perturbation from that equilibrium value. We then proceed by linearizing equations \eqref{eq:SQHD-a} and \eqref{eq:SQHD-b}, assuming that the velocity $\boldsymbol{v}$ and the non-equilibrium density $\delta\rho$ are small quantities. This leads to
\begin{align}
	\frac{\partial\delta\rho}{\partial t} & \approx-\boldsymbol{\nabla}\cdot\left[\rho_{\text{eq}}\boldsymbol{v}\right]+\sqrt{-i\frac{g}{\hbar}}\rho_{\text{eq}}\,(\xi_{1}+i\xi_{2}),\label{eq:SLL1-a}\\
	\frac{\partial\boldsymbol{v}}{\partial t} & \approx-\boldsymbol{\nabla}\left[ \frac{\delta\tilde{\mu}}{m}+\frac{1}{2m}\sqrt{i\hbar g}\,(\xi_{1}-i\xi_{2})\right],\label{eq:SLL1-b}
\end{align}
where
\begin{align}
	\delta\tilde{\mu} = g \delta \rho - \frac{\hbar^{2}}{2m}\left[\frac{1}{2\sqrt{\rho_{\text{eq}}}}\laplacian \frac{\delta \rho}{\sqrt{\rho_{\text{eq}}}} - \frac{\delta \rho}{2\rho_{\text{eq}}^{3/2}}\laplacian \sqrt{\rho_{\text{eq}}}\right]
\end{align}
comes from linearizing $\tilde{\mu}=V_{\text{ext}} + g\rho - \frac{\hbar^{2}}{2m \sqrt{\rho}} \laplacian \sqrt{\rho}$.

Taking the expectation value of these equations results in
\begin{align}
	\frac{\partial\expval{\delta\rho}}{\partial t} & \approx-\boldsymbol{\nabla}\cdot\left[\rho_{\text{eq}}\expval{\boldsymbol{v}}\right],\label{eq:LQHD-a}\\
	\frac{\partial\expval{\boldsymbol{v}}}{\partial t} & \approx-\boldsymbol{\nabla}\left[ \frac{\expval{\delta\tilde{\mu}}}{m}\right],\label{eq:LQHD-b}
\end{align}
which are precisely the equations of motion one obtains when considering low-energy excitations on top of the superfluid hydrodynamic equations \eqref{eq:QHD-a}--\eqref{eq:QHD-c} \cite{Pethick&Smith2008}. Computing the time derivative of \eqref{eq:LQHD-a} and eliminating the velocity using \eqref{eq:LQHD-b} gives
\begin{align}
	\frac{\partial^{2}\expval{\delta\rho}}{\partial t^{2}} & \approx\boldsymbol{\nabla}\cdot\left[\rho_{\text{eq}}\boldsymbol{\nabla}\left(\frac{\expval{\delta\tilde{\mu}}}{m}\right)\right],
\end{align}
which governs the dynamics of elementary excitations of Bose gases in arbitrary potentials, and can be used to derive the Bogoliubov excitation spectrum as well as the frequencies of collective oscillations in trapped gases \cite{Pethick&Smith2008,Menotti2002}. 

Returning now to equations \eqref{eq:SLL1-a} and \eqref{eq:SLL1-b}, we consider the case of a uniform box where $\rho_{\text{eq}}=$ constant, resulting in $\delta\tilde{\mu}=g\delta\rho-\frac{\hbar^{2}}{4m}\frac{1}{\rho_{\text{eq}}}\laplacian\delta\rho$. Then, ignoring the effects of quantum pressure, the stochastic equations of motion are reduced to
\begin{align}
	\frac{\partial\delta\rho}{\partial t} & \approx-\boldsymbol{\nabla}\cdot\left[\rho_{\text{eq}}\boldsymbol{v}\right]+\sqrt{-i\frac{g}{\hbar}}\rho_{\text{eq}}\,(\xi_{1}+i\xi_{2}),\label{eq:SLL3D-a}\\
	\frac{\partial\boldsymbol{v}}{\partial t} & \approx-\boldsymbol{\nabla}\left[ \frac{g\delta\rho}{m}+\frac{1}{2m}\sqrt{i\hbar g}\,(\xi_{1}-i\xi_{2})\right].\label{eq:SLL3D-b}
\end{align}
Since it is the quantum pressure term(s) which contribute to the quadratic part of the Bogoliubov spectrum, having ignored them restricts our consideration to only the linear part of the spectrum. We point out however that these equations of motion go beyond the usual superfluid description due to their stochastic nature, and by virtue of the noise terms they can be used to compute the non-equilibrium correlation functions of low-energy excitations.

When considering the one-dimensional regime, equations \eqref{eq:SLL3D-a} and \eqref{eq:SLL3D-b} give
\begin{align}
	\frac{\partial\delta\rho}{\partial t} & \approx-\frac{\partial}{\partial x}\left[\rho_{\text{eq}}v\right]+\sqrt{-i\frac{g}{\hbar}}\rho_{\text{eq}}\,(\xi_{1}+i\xi_{2}),\label{eq:SLL-a}\\
	\frac{\partial v}{\partial t} & \approx-\frac{\partial}{\partial x}\left[ \frac{g\delta\rho}{m}+\frac{1}{2m}\sqrt{i\hbar g}\,(\xi_{1}-i\xi_{2})\right],\label{eq:SLL-b}
\end{align}
which we call the stochastic Luttinger liquid (SLL) equations for a weakly-interacting Bose gas since they represent a stochastic version of the usual Luttinger liquid equations of motion in the respective operator form \cite{Cazalilla2004,Cazalilla2011},
\begin{align}
	\frac{\partial \delta \hat{\rho}}{\partial t} &= - \frac{\partial}{\partial x} \left[\rho_{\text{eq}} \hat{v}\right], \\
	\frac{\partial \hat{v}}{\partial t} &= - \frac{\partial}{\partial x} \left[\frac{g}{m} \delta \hat{\rho}\right],
\end{align}
with the understanding that quantum fluctuations in the SLL equations are incorporated via the stochastic noise terms.

We mention also that the SLL equations can be derived directly from the LSQHD equations \eqref{eq:LSQHD-a} and \eqref{eq:LSQHD-b} by assuming that the  deterministic mean-field variables are static (do not change in time), and then more specifically that they describe a homogeneous gas, along with ignoring any terms that resulted from the many-body quantum pressure. This corresponds formally to taking: (\emph{i}) $\rho_{0}=\rho_{\text{eq}}$, (\emph{ii}) $\boldsymbol{v}_{0}=0$ with the understanding that now $\delta \boldsymbol{v}\leftrightarrow\boldsymbol{v}$, and (\emph{iii}) ignoring the terms $\frac{\hbar^{2}}{4m^{2}\sqrt{\rho_{0}}}\laplacian\left(\frac{\delta\rho}{\sqrt{\rho_{0}}}\right)$ and $\frac{\hbar^{2}}{4m^{2}}\frac{\delta\rho}{\rho_{0}^{3/2}}\laplacian\sqrt{\rho_{0}}$ from equation \eqref{eq:LSQHD-b}.

\section{Physical observables} \label{sec:Physics observables}

\subsection{In the full stochastic quantum hydrodynamics} \label{sec: In the full stochastic quantum hydrodynamics}
Here we consider a few typical physical observables and reformulate them within the SQHD approach using equations \eqref{eq:fields-density-phase1} and \eqref{eq:fields-density-phase2}. We recall that observables described by averages over normally-ordered products of field operators can be computed from their stochastic counterparts using Eq. \eqref{eq:correspondence}.

Then, in this formulation, the reduced one-body density matrix and the particle number density are given by
\begin{align}
	G^{(1)}(\mathbf{r},\mathbf{r}^{\prime},t) & \equiv\langle\hat{\Psi}^{\dagger}(\mathbf{r},t)\hat{\Psi}(\mathbf{r}^{\prime},t)\rangle=\langle\widetilde{\Psi}(\mathbf{r},t)\Psi(\mathbf{r}^{\prime},t)\rangle\nonumber \\
	& =\langle\sqrt{\rho(\mathbf{r},t)\rho(\mathbf{r}^{\prime},t)}\, e^{-i\left(S(\mathbf{r},t)-S(\mathbf{r}^{\prime},t)\right)^{\,}}\rangle
\end{align}
and
\begin{align}
	G^{(1)}(\mathbf{r},\mathbf{r},t) & \equiv\langle\hat{\Psi}^{\dagger}(\mathbf{r},t)\hat{\Psi}(\mathbf{r},t)\rangle=\langle\widetilde{\Psi}(\mathbf{r},t)\Psi(\mathbf{r},t)\rangle\nonumber \\
	&=\left\langle\rho(\mathbf{r},t)\right\rangle,
\end{align}
respectively. The reduced one-body density matrix can then be normalized to obtain
\begin{align}
	g^{(1)}(\mathbf{r},\mathbf{r}^{\prime},t)&\equiv\frac{G^{(1)}(\mathbf{r},\mathbf{r}^{\prime},t)}{\sqrt{G^{(1)}(\mathbf{r},\mathbf{r},t)}\sqrt{G^{(1)}(\mathbf{r}^{\prime},\mathbf{r}^{\prime},t)}}\nonumber\\
	&=\frac{\langle\sqrt{\rho(\mathbf{r},t)\rho(\mathbf{r}^{\prime},t)}\, e^{-i\left(S(\mathbf{r},t)-S(\mathbf{r}^{\prime},t)\right)^{\,}}\rangle}{\sqrt{\left\langle\rho(\mathbf{r},t)\right\rangle}\sqrt{\left\langle\rho(\mathbf{r}^{\prime},t)\right\rangle}}. \label{eq:g1_full-hydro}
\end{align}

Furthermore, the density-density correlation function (or the reduced two-body density matrix) in normally-ordered form can be computed using
\begin{align}
	G^{(2)}(\mathbf{r},\mathbf{r}^{\prime},t) & \equiv\langle\hat{\Psi}^{\dagger}(\mathbf{r},t)\hat{\Psi}^{\dagger}(\mathbf{r}^{\prime},t)\hat{\Psi}(\mathbf{r}^{\prime},t)\hat{\Psi}(\mathbf{r},t)\rangle\nonumber \\
	& =\left\langle \rho(\mathbf{r},t)\rho(\mathbf{r}^{\prime},t)\right\rangle,
\end{align}
and normalized as
\begin{align}
	g^{(2)}(\mathbf{r},\mathbf{r}^{\prime},t)&\equiv\frac{G^{(2)}(\mathbf{r},\mathbf{r}^{\prime},t)}{G^{(1)}(\mathbf{r},\mathbf{r},t)~G^{(1)}(\mathbf{r}^{\prime},\mathbf{r}^{\prime},t)} \nonumber\\
	&=\frac{\left\langle\rho(\mathbf{r},t)\rho(\mathbf{r}^{\prime},t)\right\rangle}{\left\langle\rho(\mathbf{r},t)\right\rangle\left\langle\rho(\mathbf{r}^{\prime},t)\right\rangle}.
\end{align}
Note that, in particular, the reduced one-body density matrix \eqref{eq:g1_full-hydro} is easily represented in terms of the density and phase variables (rather than in terms of the density-velocity pair) and therefore the respective set of SDEs appears to be a more natural one to solve numerically for this observable.

\subsection{In the linearized treatment of quantum fluctuations} \label{sec:In the linearised treatment of quantum fluctuations}

In the LSQHD approach, the same quantities as above can be evaluated as follows.

The reduced one-body density matrix is given by
\begin{align}
	G&^{(1)}(\mathbf{r},\mathbf{r}^{\prime}) \nonumber\\
	&\approx \sqrt{\rho_{0}(\mathbf{r}) \rho_{0}(\mathbf{r}^{\prime})} e^{-i\left[S_{0}(\mathbf{r}) - S_{0}(\mathbf{r}^{\prime})\right]} \nonumber\\
	&\quad\times \left( 1+ \expval{\delta S(\mathbf{r})\delta S(\mathbf{r}^{\prime})} - \frac{1}{2} \left[\expval{\delta S^{2}(\mathbf{r})} + \expval{\delta S^{2}(\mathbf{r}^{\prime})}\right] \right. \nonumber\\
	&\qquad\quad + \frac{\expval{\delta \rho(\mathbf{r})\delta \rho(\mathbf{r}^{\prime})}}{4\rho_{0}(\mathbf{r})\rho_{0}(\mathbf{r}^{\prime})} - \frac{1}{8} \left[\frac{\expval{\delta\rho^{2}(\mathbf{r})}}{\rho_{0}^{2}(\mathbf{r})} + \frac{\expval{\delta\rho^{2}(\mathbf{r}^{\prime})}}{\rho_{0}^{2}(\mathbf{r}^{\prime})}\right] \nonumber\\
	&\qquad\quad + \frac{i}{2} \left[\frac{\expval{\delta \rho(\mathbf{r})\delta S(\mathbf{r}^{\prime})}}{\rho_{0}(\mathbf{r})} -  \frac{\expval{\delta S(\mathbf{r})\delta \rho(\mathbf{r}^{\prime})}}{\rho_{0}(\mathbf{r}^{\prime})}\right] \nonumber\\
	&\qquad + \left. \frac{i}{2} \left[\frac{\expval{\delta \rho(\mathbf{r}^{\prime})\delta S(\mathbf{r}^{\prime})}}{\rho_{0}(\mathbf{r}^{\prime})} - \frac{\expval{\delta S(\mathbf{r})\delta \rho(\mathbf{r})}}{\rho_{0}(\mathbf{r})}\right] \right),
\end{align}
where the time index has been dropped for brevity and we recall that averages of first order terms are zero for linearized approaches.

The particle number density itself is then simply given by
\begin{equation}
	G^{(1)}(\mathbf{r},\mathbf{r},t)=\rho_{0}(\mathbf{r},t), \label{eq:G1rr_LSQH}
\end{equation}
and normalizing the reduced one-body density matrix results in
\begin{align}
	g&^{(1)}(\mathbf{r},\mathbf{r}^{\prime})\nonumber\\
	&\approx e^{-i\left[S_{0}(\mathbf{r}) - S_{0}(\mathbf{r}^{\prime})\right]} \nonumber\\
	&\quad\times \left( 1+ \expval{\delta S(\mathbf{r})\delta S(\mathbf{r}^{\prime})} - \frac{1}{2} \left[\expval{\delta S^{2}(\mathbf{r})} + \expval{\delta S^{2}(\mathbf{r}^{\prime})}\right] \right. \nonumber\\
	& \qquad\quad + \frac{\expval{\delta \rho(\mathbf{r})\delta \rho(\mathbf{r}^{\prime})}}{4\rho_{0}(\mathbf{r})\rho_{0}(\mathbf{r}^{\prime})} - \frac{1}{8} \left[\frac{\expval{\delta\rho^{2}(\mathbf{r})}}{\rho_{0}^{2}(\mathbf{r})} + \frac{\expval{\delta\rho^{2}(\mathbf{r}^{\prime})}}{\rho_{0}^{2}(\mathbf{r}^{\prime})}\right] \nonumber\\
	&\qquad\quad + \frac{i}{2} \left[\frac{\expval{\delta \rho(\mathbf{r})\delta S(\mathbf{r}^{\prime})}}{\rho_{0}(\mathbf{r})} -  \frac{\expval{\delta S(\mathbf{r})\delta \rho(\mathbf{r}^{\prime})}}{\rho_{0}(\mathbf{r}^{\prime})}\right] \nonumber\\
	&\quad\qquad + \left. \frac{i}{2} \left[\frac{\expval{\delta \rho(\mathbf{r}^{\prime})\delta S(\mathbf{r}^{\prime})}}{\rho_{0}(\mathbf{r}^{\prime})} - \frac{\expval{\delta S(\mathbf{r})\delta \rho(\mathbf{r})}}{\rho_{0}(\mathbf{r})}\right] \right), \label{eq:g1_LSQH}
\end{align}
where we drop the time index again for brevity.

Furthermore, the density-density correlation function in normally-ordered form can be computed using
\begin{equation}
	G^{(2)}(\mathbf{r},\mathbf{r}^{\prime},t)=\rho_{0}(\mathbf{r},t)\rho_{0}(\mathbf{r}^{\prime},t)+\left\langle\delta\rho(\mathbf{r},t)\delta\rho(\mathbf{r}^{\prime},t)\right\rangle,
\end{equation}
and normalised as
\begin{align}
	g^{(2)}(\mathbf{r},\mathbf{r}^{\prime},t)&\equiv\frac{G^{(2)}(\mathbf{r},\mathbf{r}^{\prime},t)}{G^{(1)}(\mathbf{r},\mathbf{r},t)~G^{(1)}(\mathbf{r}^{\prime},\mathbf{r}^{\prime},t)} \nonumber\\
	&=1+\frac{\left\langle\delta\rho(\mathbf{r},t)\delta\rho(\mathbf{r}^{\prime},t)\right\rangle}{\rho_{0}(\mathbf{r},t)\rho_{0}(\mathbf{r}^{\prime},t)}. \label{eq:g2_LSQH}
\end{align}

In the stochastic Bogoliubov approach, on the other hand, the reduced one-body density matrix can be computed using
\begin{align}
	G^{(1)}(\mathbf{r},\mathbf{r}^{\prime},t)
	&=\Psi_{0}^{*}(\mathbf{r},t)\Psi_{0}(\mathbf{r}^{\prime},t)+\Psi_{0}^{*}(\mathbf{r},t)\langle\delta\psi(\mathbf{r}^{\prime},t)\rangle\nonumber\\
	&\qquad+\Psi_{0}(\mathbf{r}^{\prime},t)\langle \delta\tilde{\psi}(\mathbf{r},t)\rangle+\langle\delta\tilde{\psi}(\mathbf{r},t) \delta\psi(\mathbf{r}^{\prime},t)\rangle\nonumber\\
	&=\Psi_{0}^{*}(\mathbf{r},t)\Psi_{0}(\mathbf{r}^{\prime},t)+\langle\delta\tilde{\psi}(\mathbf{r},t) \delta\psi(\mathbf{r}^{\prime},t)\rangle,
\end{align}
since averages of first order terms are zero.

The particle number density itself is then given by
\begin{align}
	G^{(1)}(\mathbf{r},\mathbf{r},t)
	&=\left|\Psi_{0}(\mathbf{r},t)\right|^{2}+\langle\delta\tilde{\psi}(\mathbf{r},t) \delta\psi(\mathbf{r},t)\rangle, \label{eq:G1rr_SB}
\end{align}
and the normalized reduced one-body density matrix is (see Appendix \ref{appendix:stoch.bogo.observables})
\begin{align}
	g&^{(1)}(\mathbf{r},\mathbf{r}^{\prime},t)\nonumber\\
	&\approx
	\sqrt{\frac{\Psi_{0}^{*}(\mathbf{r},t)\Psi_{0}(\mathbf{r}^{\prime},t)}{\Psi_{0}(\mathbf{r},t) \Psi_{0}^{*}(\mathbf{r}^{\prime},t)}}
	\left(1 + \frac{ \langle\delta \tilde{\psi}(\mathbf{r},t)\delta \psi(\mathbf{r}^{\prime},t)\rangle}{\Psi_{0}^{*}(\mathbf{r},t)\Psi_{0}(\mathbf{r}^{\prime},t)} \right.\nonumber\\
	&\,\,\, - \left. \frac{1}{2}\frac{\langle\delta \tilde{\psi}(\mathbf{r},t)\delta \psi(\mathbf{r},t)\rangle}{\Psi_{0}^{*}(\mathbf{r},t)\Psi_{0}(\mathbf{r},t)} - \frac{1}{2}\frac{\langle\delta \tilde{\psi}(\mathbf{r}^{\prime},t)\delta \psi(\mathbf{r}^{\prime},t)\rangle}{\Psi_{0}^{*}(\mathbf{r}^{\prime},t)\Psi_{0}(\mathbf{r}^{\prime},t)} \right). \label{eq:g1_SB}
\end{align}

As mentioned previously in Sec. \ref{sec:Comparison with Bogoliubov approaches} regarding the real-space density, a similar discrepancy exists between the normalized reduced one-body density matrices \eqref{eq:g1_LSQH} and \eqref{eq:g1_SB}. These expressions are not equivalent under the transformations \eqref{eq:DeltaPsi_HydroTransform} and \eqref{eq:DeltaPsiTilde_HydroTransform} due to the inconsistency of maintaining second order terms in these expressions (which are needed to obtain leading order corrections to the mean-field results) while the equations of motion themselves have been truncated at linear order. A similar type of inconsistency has been identified in Ref. \cite{Castin2003} when developing an extension of Bogoliubov theory to quasi-condensates. We explore these differences between the LSQHD and stochastic Bogoliubov predictions in Appendix \ref{appendix:g1_compare}, where we show that differences in the density are noticeable, whereas differences in $g^{(1)}(\mathbf{r},\mathbf{r}^{\prime},t)$ are not.

Finally, the normalized density-density correlation function is given by (see Appendix \ref{appendix:stoch.bogo.observables})
\begin{align}
	g^{(2)}(\mathbf{r},&\mathbf{r}^{\prime},t) \approx 1+2\Re{\frac{\langle\delta\psi(\mathbf{r},t)\delta\psi(\mathbf{r}^{\prime},t)\rangle}{\Psi_{0}(\mathbf{r},t)\Psi_{0}(\mathbf{r}^{\prime},t)}}\nonumber\\
	&+\frac{\langle\delta\tilde{\psi}(\mathbf{r},t)\delta\psi(\mathbf{r}^{\prime},t)\rangle}{\Psi_{0}^{*}(\mathbf{r},t)\Psi_{0}(\mathbf{r}^{\prime},t)}+\frac{\langle\delta\tilde{\psi}(\mathbf{r}^{\prime},t)\delta\psi(\mathbf{r},t)\rangle}{\Psi_{0}^{*}(\mathbf{r}^{\prime},t)\Psi_{0}(\mathbf{r},t)}. \label{eq:g2_SB}
\end{align}

We point out that whilst the expressions for the normalized first order correlators \eqref{eq:g1_LSQH} and \eqref{eq:g1_SB} are inconsistent, the normalized second order correlators \eqref{eq:g2_LSQH} and \eqref{eq:g2_SB} are in fact equivalent under the transformations \eqref{eq:DeltaPsi_HydroTransform} and \eqref{eq:DeltaPsiTilde_HydroTransform}.

\section{Simulations \& Comparisons with existing methods}

\subsection{Density-density correlations in quantum shock waves} \label{sec:Density-density correlations in quantum shock waves}
As a first application of the SQHD formalism to systems of interacting bosons, we begin by exploring the correlations arising in a scenario which develops dispersive quantum shock waves. In particular we focus on the set-up previously investigated in Ref. \cite{Simmons2020}: namely, an initial density bump on top of a uniform non-zero background in a one-dimensional (1D) Bose gas. Considering two examples with different particle number $N$, yet with identical 1D interaction strength $g_{\text{1D}}$, we compare the predictions made by SQHD with other well-known stochastic methods, as well as exact quantum calculations using matrix product states (MPS).

In each example we consider an initial density profile of the form $G^{(1)}(x,x,0)=\rho(x,0)=N_{\text{bg}}[1+\beta e^{-x^{2}/2\sigma^2}]^{2}/L$ which subsequently evolves in a uniform potential of length $L$ with periodic boundary conditions. Here $N_{\text{bg}}=\rho_{\text{bg}}L$ specifies the number of particles in the background, related to the total particle number via $N=N_{\text{bg}}(1+\frac{\sqrt{\pi}\beta\sigma}{L}[\beta\erf(\frac{L}{2\sigma})+2\sqrt{2}\erf(\frac{L}{2\sqrt{2}\sigma})])$. The amplitude of the bump above the background density is determined by $\beta$ and its width is governed by $\sigma$.

For this scenario one can define a local dimensionless interaction parameter $\gamma(x)=mg_{\text{1D}}/\hbar^{2}\rho(x)$ (or a local Lieb-Liniger parameter \cite{LiebLiniger1963,Kheruntsyan_2005}), and then use its value at the background density $\gamma_{\text{bg}}=mg_{\text{1D}}/\hbar^{2}\rho_{\text{bg}}$ to characterize the initial state. Such a parameter then allows for a simple definition of the characteristic length-scale associated with interactions---the dimensionless healing length $l_{\text{h}}/L=1/N_{\text{bg}}\sqrt{\gamma_{\text{bg}}}$, where we use $L$ as the length scale.

\subsubsection{Example 1: $\sigma/l_{h}\simeq0.11$\\($N=50$, $\gamma_{\text{bg}}=0.1$, $\bar{g}_{\text{1D}}\simeq4.773$)} \label{sec:Example1}

In this first example we consider a weakly-interacting 1D Bose gas of $N=50$ particles, with $\gamma_{\text{bg}}=0.1$. The initial bump is characterized by $\beta=1$ and $\sigma/L=0.007$. This configuration leads to $N_{\text{bg}}\simeq47.73$ particles in the background, and therefore a background healing length of $l_\text{h}/L\simeq0.066$. As such, the bump width dominates the dynamics as the shortest length-scale in the problem \cite{Simmons2020}, being only about 10\% of the background healing length, $\sigma\simeq0.11l_{\text{h}}$. For completeness, the dimensionless interaction strength for this scenario is given by $\bar{g}_{\text{1D}}=g_{\text{1D}}mL/\hbar^{2}=\gamma_{\text{bg}}N_{\text{bg}}\simeq4.773$.

In the following figures we show predictions for the dimensionless real-space density $G^{(1)}(x,x,t)L=\rho(x,t)L$ and the normally-ordered density-density correlation function $g^{(2)}(x,x^{\prime},t)$ that result from a number of different quantum many-body approaches. These include; the full stochastic positive-\textit{P} equations, the LSQHD equations, and the truncated Wigner approach (see Appendix \ref{appendix:Wigner} for details), along with an infinite-MPS calculation (see Appendix \ref{appendix:MPS} for details). We have also simulated the stochastic Bogoliubov equations, but have omitted these results from the figures below for clarity since we recall that the respective density-density correlation function $g^{(2)}(x,x^{\prime},t)$ is equivalent to that of LSQHD, whereas results for the predicted real-space density can be found in Appendix \ref{appendix:g1_compare}.

\begin{figure}[tbp]
	\centering
	\includegraphics[width=1.0\linewidth]{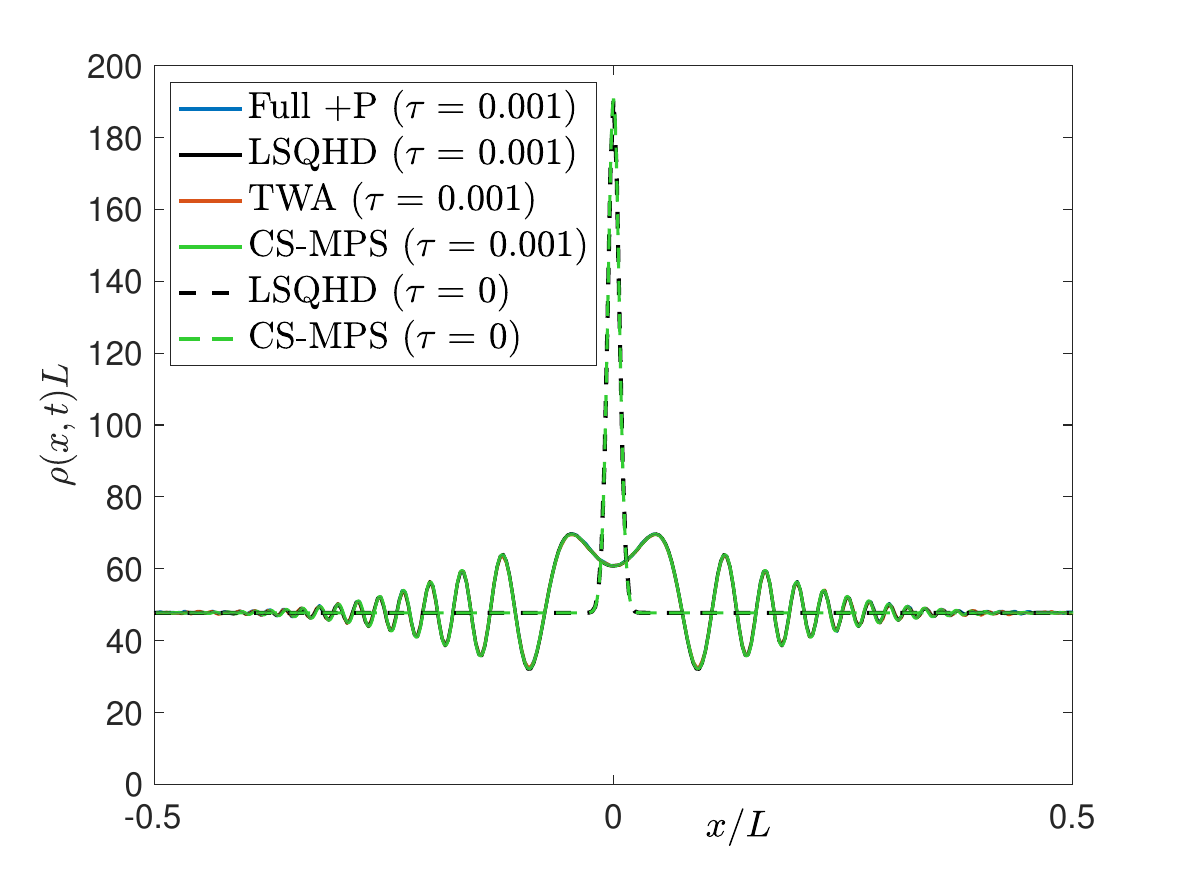}
	\caption{The real-space density $\rho(x,t)$ of a weakly interacting 1D Bose gas in the quantum shock wave scenario of Ref. \cite{Simmons2020}. Shown are the initial ($\tau=0$) and time-evolved ($\tau=0.001$) dimensionless density profiles $\rho(x,t)L$, where the dimensionless time $\tau$ is introduced according to $\tau=t\hbar/mL^{2}$, for $N=50$ particles and a dimensionless background interaction strength of $\gamma_{\text{bg}}=0.1$. The initial state (dashed lines) has a width $\sigma/L=0.007$ and amplitude $\beta=1$. This configuration leads to $N_{\text{bg}}\simeq47.73$ particles in the background, and a background healing length of $l_{\text{h}}/L\simeq0.066$. The results displayed are for: the full stochastic positive-\textit{P} equations (Full +P), the linearized stochastic quantum hydrodynamic equations (LSQHD), the truncated Wigner approach (TWA), and a matrix product state calculation initialized in a coherent state (CS-MPS), with all the different curves being essentially indistinguishable from each other on this scale. Other technical details on simulations are provided in the caption to Table
	\ref{table:data_N50_gammabg0.1_sigmabar0.007} below.
	}
	\label{fig:N50_gammabg0.1_sigmabar0.007-density_comparison}
\end{figure}

Each of the approaches used here have been initialized in a coherent state (see Appendices \ref{appendix:Wigner}, \ref{appendix:initial_states} and \ref{appendix:MPS}) for a fair comparison with the positive-\textit{P} schemes, since a more suitable approximation to the initial many-body ground state within this formalism does not currently exist in the literature. Out of interest, in Appendix \ref{appendix:GS-MPS_compare} we compare these initial coherent state results with an infinite-MPS calculation initialized in the exact many-body ground state of a trapping potential which is tailored to closely reproduce the same initial density  $\rho(x,0)=N_{\text{bg}}[1+\beta e^{-x^{2}/2\sigma^2}]^{2}/L$ as the mean-field (coherent state) approach.

\begin{figure}[htbp]
	\centering
	\includegraphics[width=1.0\linewidth]{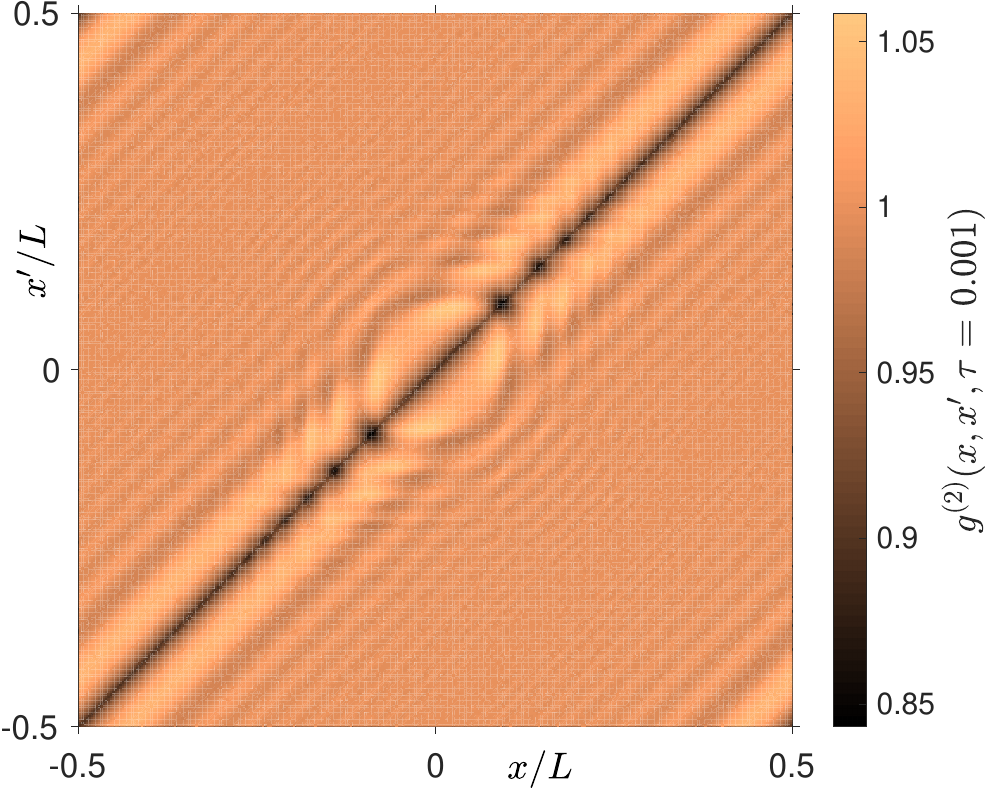}
	\caption{The density-density correlation function $g^{(2)}(x,x^{\prime},t)$ from LSQHD at dimensionless time $\tau=0.001$ for the same parameters as in Fig. \ref{fig:N50_gammabg0.1_sigmabar0.007-density_comparison}.}
	\label{fig:N50_gammabg0.1_sigmabar0.007-g2xxp_comparison}
\end{figure}

In Figure \ref{fig:N50_gammabg0.1_sigmabar0.007-density_comparison} we show the real-space density at the initial dimensionless time $\tau=t\hbar/mL^{2}=0$ and at time $\tau=0.001$. When released into the uniform trap, the initial density bump begins to split into left and right moving parts, where each quickly develops into a dispersive quantum shock wave---an oscillatory wave-train that results from quantum mechanical self-interference \cite{Simmons2020}. Unlike each of the other methods presented here, we recall that the LSQHD scheme does not incorporate the effect of quantum fluctuations into the density, \textit{i.e.} it reproduces only the mean-field superfluid result for this observable. Nevertheless the densities predicted by each method in Fig. \ref{fig:N50_gammabg0.1_sigmabar0.007-density_comparison} agree very well, and on the scale plotted here they lie almost directly on top of each other making them virtually indistinguishable. Hence, the effect of quantum fluctuations on the real-space density (which is known to reduce the amplitude of the interference contrast from that of the pure mean-field GPE prediction \cite{Simmons2020}) is negligible in this scenario. Their limited effect on the density is discussed in Appendix \ref{appendix:GS-MPS_compare}.

Turning now to examine the quantity of most interest in this section, in Figure \ref{fig:N50_gammabg0.1_sigmabar0.007-g2xxp_comparison} we plot the entire normalized density-density correlation function $g^{(2)}(x,x^{\prime},t)$ at time $\tau=0.001$, as predicted by LSQHD. We then proceed by exploring and comparing relevant slices of this correlation function, demonstrating the advantages and usefulness of the LSQHD approach.

\begin{figure}[htbp]
	\centering
	\includegraphics[width=1.0\linewidth]{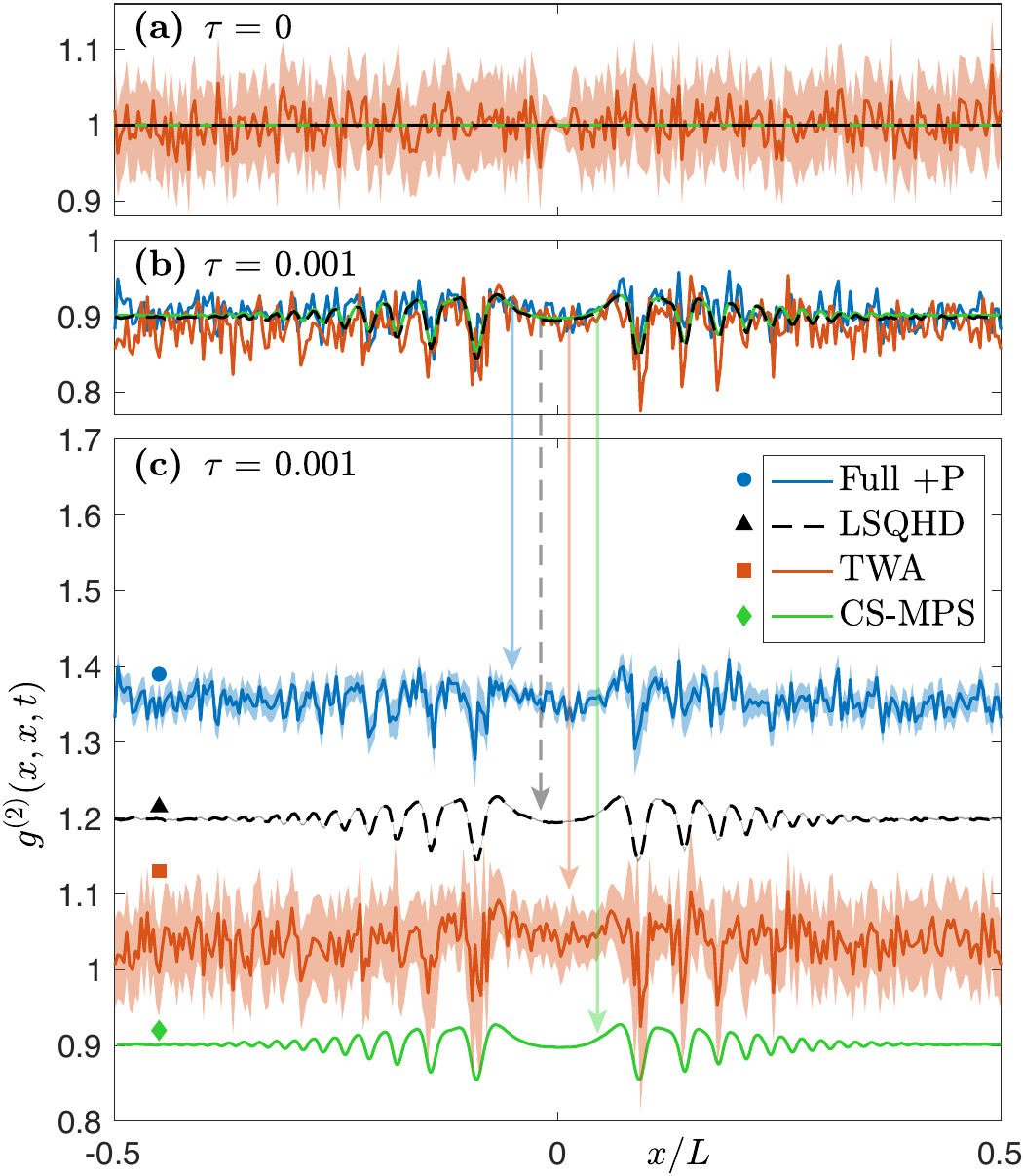}
	\caption{The same-point correlation function $g^{(2)}(x,x,t)$ for the scenario presented in Fig. \ref{fig:N50_gammabg0.1_sigmabar0.007-density_comparison}. (a) shows the initial correlations at $\tau=0$ which are precisely $g^{(2)}(x,x,0)=1$ for each method, except for the truncated Wigner approach (TWA) which contains initial fluctuations around this value that go to zero in the limit of infinitely many stochastic trajectories. (b) provides a comparison of $g^{(2)}(x,x,\tau=0.001)$ for each method (without any uncertainty shown). For clarity, (c) shows the same data as in (b) spaced upward by 0.15 from the coherent-state MPS (CS-MPS) result, and arranged so that the order of the figure legend coincides with the vertical ordering of the results (as indicated by the corresponding marker symbols). The shaded regions in (a) and (c) denote one standard error of uncertainty for the respective stochastic approaches, which for linearized stochastic quantum hydrodynamics (LSQHD) grows dynamically to about half the size of the linewidth used here and can be seen between the dashes in (c).}
	\label{fig:N50_gammabg0.1_sigmabar0.007-g2xx_comparison}
\end{figure}

\begin{figure}[htbp]
	\centering
	\includegraphics[width=1.0\linewidth]{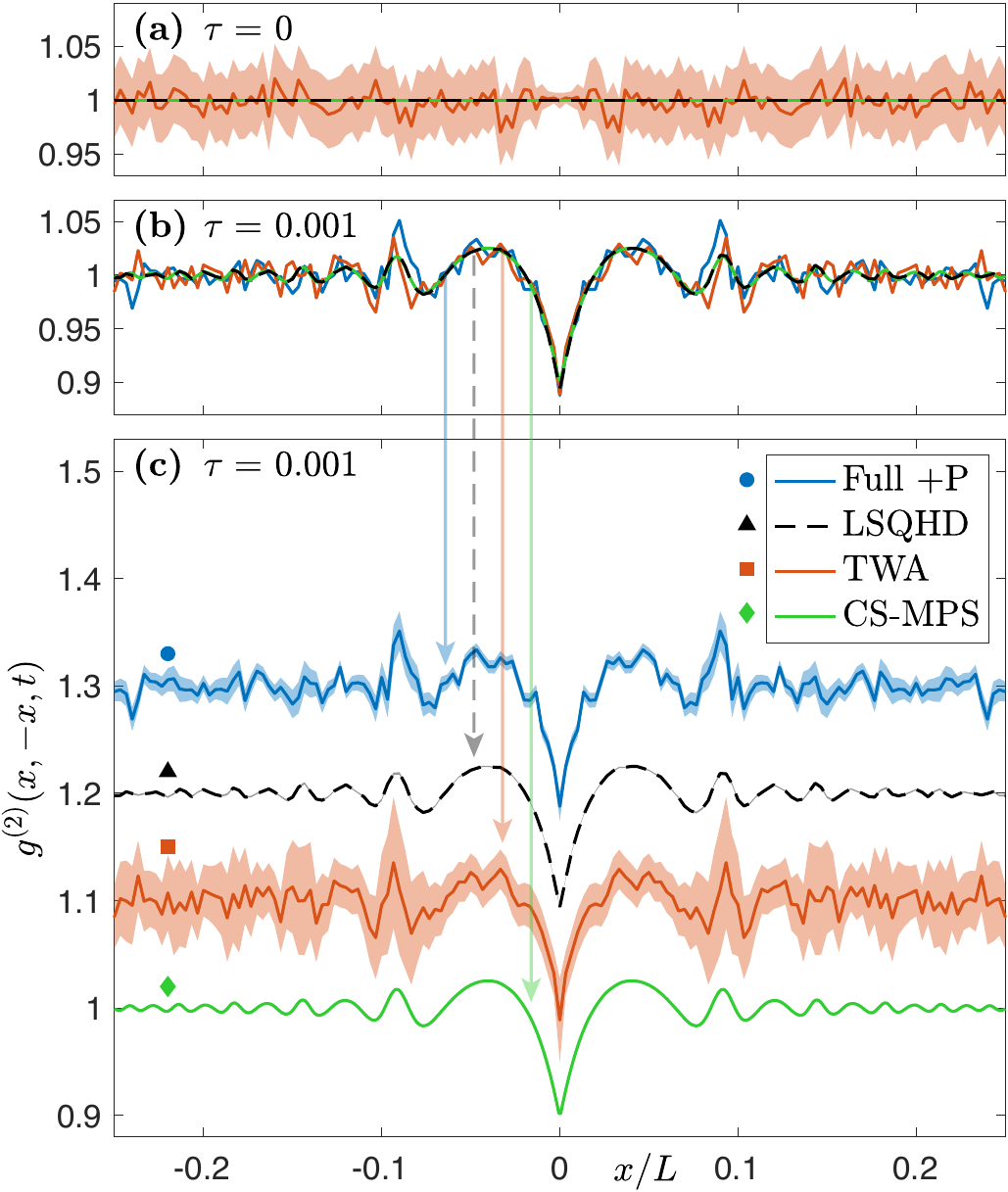}
	\caption{The opposite-point correlation function $g^{(2)}(x,-x,t)$ for the scenario presented in Fig. \ref{fig:N50_gammabg0.1_sigmabar0.007-density_comparison}. To examine the correlations resulting from the density bump itself we plot between $|x|/L=0.25$. Due to the periodic boundary conditions of the system, values outside of this domain correspond to probing the opposite-point correlations of the background. The results are presented similarly to Fig. \ref{fig:N50_gammabg0.1_sigmabar0.007-g2xx_comparison} where the upward shift in (c) is now only 0.1.}
	\label{fig:N50_gammabg0.1_sigmabar0.007-g2xmx_comparison}
\end{figure}

In Figures \ref{fig:N50_gammabg0.1_sigmabar0.007-g2xx_comparison} and \ref{fig:N50_gammabg0.1_sigmabar0.007-g2xmx_comparison}, for each of the aforementioned approaches, we plot the same-point $g^{(2)}(x,x,t)$ and opposite-point $g^{(2)}(x,-x,t)$ correlation functions respectively. These two correlation functions develop oscillations during evolution which coincide with those in the real-space density, where $g^{(2)}(x,-x,t)$ oscillates twice for every oscillation in the density. Furthermore, in Table \ref{table:data_N50_gammabg0.1_sigmabar0.007} we show the time taken to simulate this scenario for each approach, where identical lattice spacing and time stepping was used for each of the stochastic schemes.

The first advantage of the LSQHD approach here is readily apparent in the same- and opposite-point correlations: it provides sufficiently smooth results that lie close to the exact coherent-state MPS (CS-MPS) predictions, and it does not suffer from excessive noise like the full positive-\textit{P} and truncated Wigner approaches. More specifically, the LSQHD approach possesses noise that is $\approx \!17$ times smaller in magnitude compared to the full positive-\textit{P} approach, and $\approx \!54$ times smaller than the noise in the truncated Wigner result. This means that reducing the Full +P and TWA error bars down to the same level as for LSQHD would require running $\sim \!300$ and $\sim \!3000$ times more stochastic trajectories for each respective approach.

The second advantage we mention is the reduced simulation time compared with CS-MPS. The LSQHD simulation was approximately $3.3$ times faster than the exact CS-MPS calculation and yet provides essentially the same results for each of the correlation functions. Moreover, the LSQHD scheme was similarly faster to simulate than the stochastic Bogoliubov approach. While the $g^{(2)}(x,x^{\prime},t)$ results of the stochastic Bogoliubov equations lie directly underneath the LSQHD result (within one standard error), they took $\sim1.5$ times longer to simulate. For clarity, we report that no elaborate code optimization was performed for any of the stochastic simulations reported in Table \ref{table:data_N50_gammabg0.1_sigmabar0.007} and that they were each carried out on an Apple iMac desktop hosting a 4.2GHz Intel Core i7 (i7-7700K) processor with 4 cores and 64GB of RAM, whereas the MPS simulations were performed using the Matrix Product Toolkit of Ref. \cite{MPS_toolkit} on an Apple iMac desktop hosting a 3.5GHz Intel Core i7 (i7-4771) processor with 4 cores and 32GB of RAM. %\cite{Ians_Machine}.

Finally, we examine the applicability of the linearized positive-\textit{P} schemes and the truncated Wigner approach for this scenario. A check of the population growth present in the stochastic Bogoliubov scheme reveals an increase in particle number of $\approx2\%$ by the final time of the simulation, confirming the validity of the undepleted pump approximation and the reasonableness of using the linearized positive-\textit{P} schemes here. The usual requirement for applicability of the truncated Wigner approach, on the other hand, is that the number of particles $N$ be much larger than the number of modes $M$, \textit{i.e.} the lattice mode occupation should be $N/M\gg1$ \cite{Sinatra2002}. However in this case it is only $N/M=50/300\simeq0.17$ and hence the \textit{a priori} applicability of this approach is questionable. Nevertheless, we note that it still provides reasonable predictions for the density and $g^{(2)}(x,x^{\prime},t)$ (even though it suffers from excessive noise). This supports the notion that the condition $N/M\gg1$ can be somewhat relaxed in 1D to $N/M\approx1$ \cite{Ruostekoski2005,Ruostekoski2006,Ruostekoski2012}, or even a smaller value as is the case here.

\begin{table}[tbp]
	\begin{center}
		\caption{Simulation times for the results presented in Figs. \ref{fig:N50_gammabg0.1_sigmabar0.007-density_comparison}-\ref{fig:N50_gammabg0.1_sigmabar0.007-g2xmx_comparison}. The stochastic simulations were run using the XMDS software of Ref. \cite{xmds}. For each stochastic approach the results we present here are an average over 2,000,000 stochastic trajectories which were run in parallel using the ``mpi-multi-path" driver, with 300 spatial lattice points and a time step of $\Delta\tau=3.3333\times10^{-6}$. Simulation outputs were $\rho(x,t)$ and $g^{(2)}(x,x^{\prime},t)$, along with their associated standard errors. The positive-\textit{P} schemes were integrated using the SI (semi-implicit) algorithm \cite{Drummond_SI_1983,Werner_SI_1997} and the truncated Wigner scheme was integrated using the RK4 (fourth-order Runge-Kutta) algorithm.}
		\label{table:data_N50_gammabg0.1_sigmabar0.007}
		\def\arraystretch{1.3}
		\setlength{\tabcolsep}{22pt}
		\begin{tabular}{cc}
			\hline\hline
			Method & Time to simulate \\
			\hline
			Full $+P$ & 36 hours \\ 
			LSQHD & 37 hours \\ 
			Stoch. Bogoliubov & 54 hours \\ 
			TWA & 39 hours \\ 
			CS-MPS & 124 hours \\ 
			\hline\hline
		\end{tabular}
	\end{center}
\end{table}

\subsubsection{Example 2: $\sigma/l_{h}\simeq0.33$\\($N=500$, $\gamma_{\text{bg}}=0.01$, $\bar{g}_{\text{1D}}\simeq4.773$)} \label{sec:Example2}
In this second example, similarly to the first, we also consider an initial bump characterized by $\beta=1$ and $\sigma/L=0.007$. Here however, we increase the total particle number by an order of magnitude to $N=500$, while maintaining an equal dimensionless interaction strength $\bar{g}_{\text{1D}}=g_{\text{1D}}mL/\hbar^{2}=\gamma_{\text{bg}}N_{\text{bg}}\simeq4.773$. Such a configuration has a background interaction parameter of $\gamma_{\text{bg}}=0.01$ and $N_{\text{bg}}\simeq477.3$ particles in the homogeneous background. This leads to a shorter background healing length of $l_{\text{h}}/L=1/N_{\text{bg}}\sqrt{\gamma_{\text{bg}}}\simeq0.021$, where the initial bump width is now about 30\% of the background healing length, $\sigma\simeq0.33l_{\text{h}}$. Whilst only a small increase in the comparability of these two length scales, the increased role of interactions in this scenario brings about a clear qualitative change in the correlations of the system, as we will discuss in a moment.

The usefulness of the LSQHD approach is illustrated in this scenario in that it is the only one of a few quantum many-body approaches that remains tractable. The MPS approach becomes intractable here due to the large number of particles $N=500$, and the multiplicative nature of the noise in the full positive-\textit{P} treatment causes this approach to break down halfway through the full simulation time window, at $\tau_{{break}}=t_{{break}}\hbar/mL^{2}=0.0005$. This leaves only the linearized stochastic schemes and the truncated Wigner approach, each of which are initialized in a coherent state with 300 lattice points, and averages taken over 500,000 trajectories. Here we present a comparison of these methods at the final time of the simulation $\tau=0.001$.

\begin{figure}[tbp]
	\centering
	\includegraphics[width=1.0\linewidth]{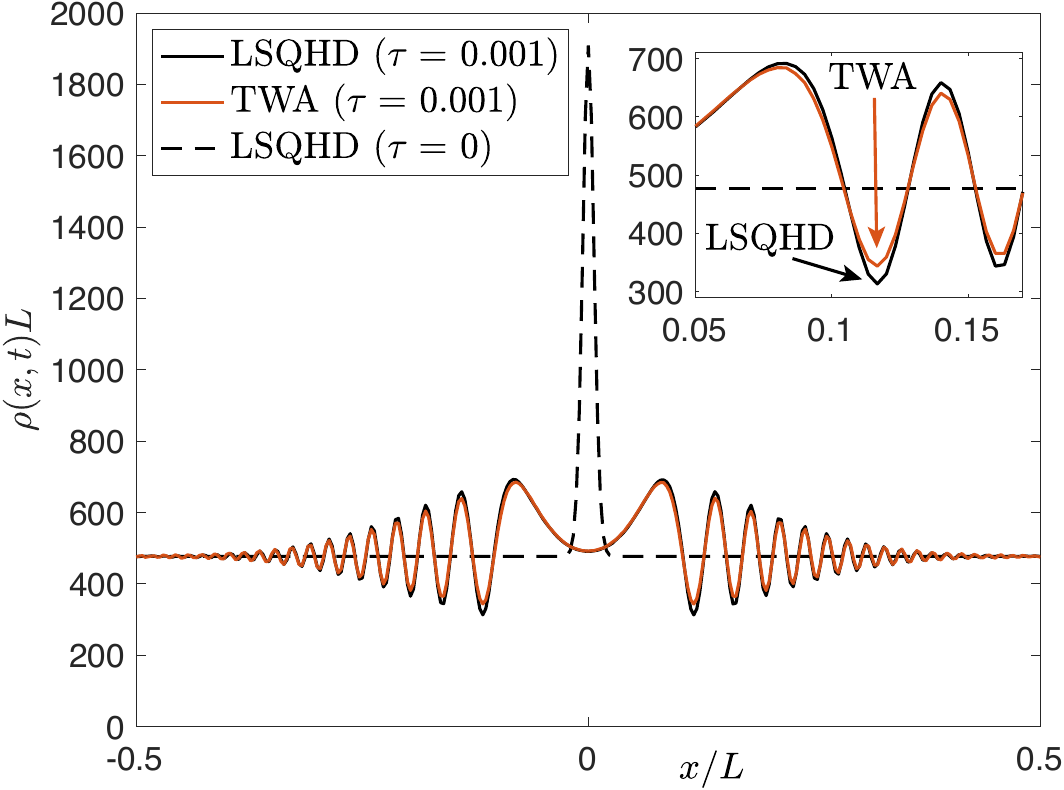}
	\caption{Density profiles in a quantum shock wave scenario as in Fig.~ \ref{fig:N50_gammabg0.1_sigmabar0.007-density_comparison}, but for $N=500$ particles and a dimensionless background interaction strength of $\gamma_{\text{bg}}=0.01$. These parameters lead to $N_{\text{bg}}\simeq477.3$, and a background healing length of $l_{\text{h}}/L\simeq0.021$. The results displayed are for the LSQHD and TWA approaches. The inset shows more clearly the effect of quantum fluctuations on the first few interference fringes; we recall that quantum fluctuations are accounted for in the TWA result, but are absent in the density of LSQHD such that $\rho(x,t)$ is given simply by the mean-field prediction $\rho(x,t)=\rho_0(x,t)=|\psi_0(x,t)|^2$ of superfluid hydrodynamics [see Eq.~\eqref{eq:G1rr_LSQH}].}
	\label{fig:N500_gammabg0.01_sigmabar0.007-density_comparison_tend}
\end{figure}

In Figure \ref{fig:N500_gammabg0.01_sigmabar0.007-density_comparison_tend} we show the evolution of the real-space density. The result is similar as in Example 1: the initial bump splits into left and right moving parts which quickly develop into dispersive quantum shock waves. Again, the density predicted by each method agrees well with each other. In contrast to Example 1 however, by the time $\tau=0.001$, we see that the left and right moving parts have almost completely separated, and that the effect of quantum fluctuations on the density is already becoming apparent on the scale plotted here -- the amplitude of oscillations in the truncated Wigner result is slightly reduced compared to the LSQHD scheme, which reproduces the mean-field density. This reduced interference contrast can be seen more clearly in the inset of Fig. \ref{fig:N500_gammabg0.01_sigmabar0.007-density_comparison_tend}.

\begin{figure}[htbp]
	\centering
	\includegraphics[width=1.0\linewidth]{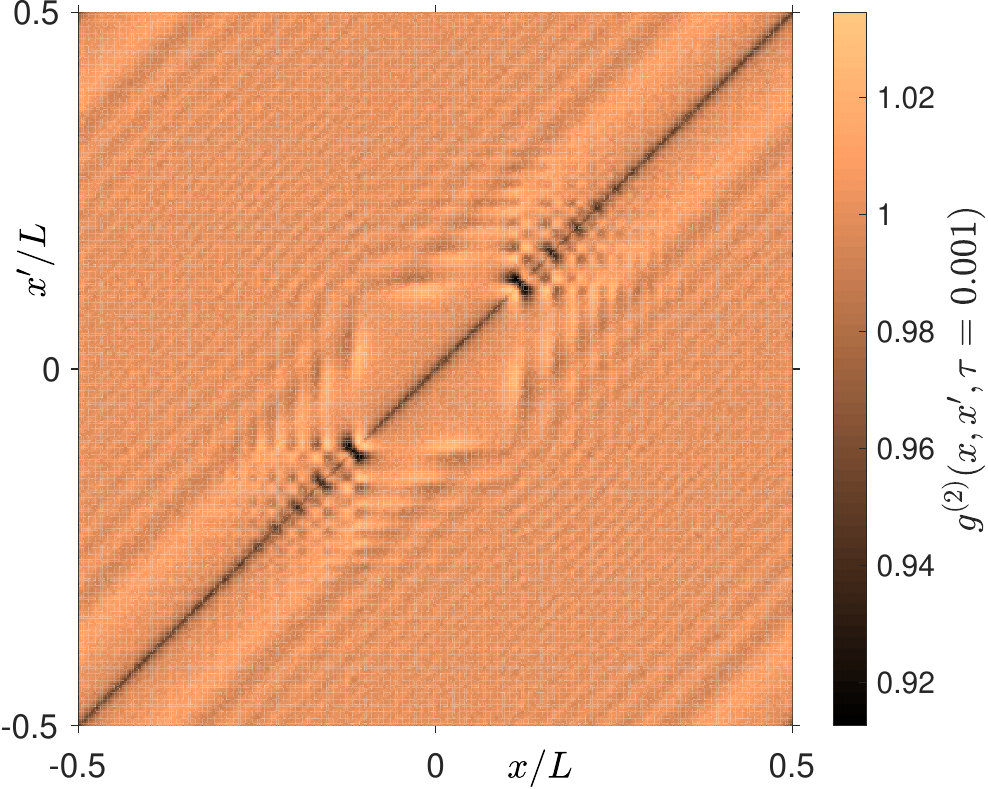}
	\caption{The density-density correlation function $g^{(2)}(x,x^{\prime},t)$ from LSQHD at dimensionless time $\tau=0.001$ for the same parameters as in Figure \ref{fig:N500_gammabg0.01_sigmabar0.007-density_comparison_tend}.}
	\label{fig:N500_gammabg0.01_sigmabar0.007-g2xxp_comparison_tend}
\end{figure}

In Figure \ref{fig:N500_gammabg0.01_sigmabar0.007-g2xxp_comparison_tend}, we plot the entire normalized density-density correlation function $g^{(2)}(x,x^{\prime},t)$ at time $\tau=0.001$, as predicted by LSQHD. Similarly, we proceed by examining the same correlation slices as before, comparing the linearized stochastic schemes with the truncated Wigner approach and further demonstrating the utility of the LSQHD equations.

\begin{figure}[htbp]
	\centering
	\includegraphics[width=1.0\linewidth]{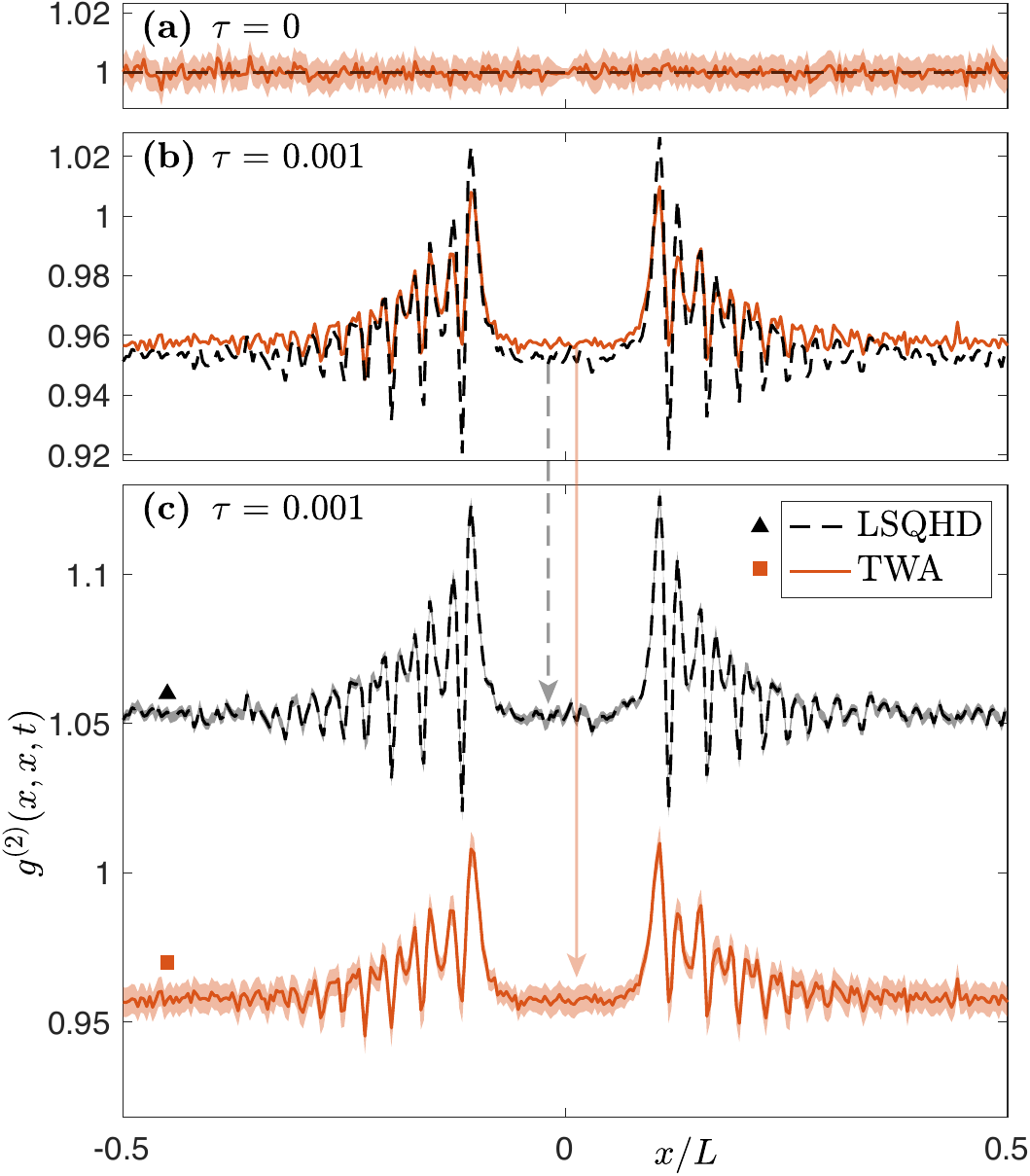}
	\caption{The same-point correlation function $g^{(2)}(x,x,t)$ as in Fig.~\ref{fig:N50_gammabg0.1_sigmabar0.007-g2xx_comparison}, but for the parameters of Fig.~\ref{fig:N500_gammabg0.01_sigmabar0.007-density_comparison_tend}. The only computationally tractable approaches in this regime are the LSQHD and TWA schemes (see text). For clarity, (c) shows the same data as in (b) spaced upward by 0.1 from the TWA result. The shaded regions in (a) and (c) denote one standard error of uncertainty for the respective stochastic approaches, which for LSQHD grows dynamically to about the size of the linewidth used here and can be seen between the dashes in (c).}
	\label{fig:N500_gammabg0.01_sigmabar0.007-g2xx_comparison_tend}
\end{figure}

\begin{figure}[htbp]
	\centering
	\includegraphics[width=1.0\linewidth]{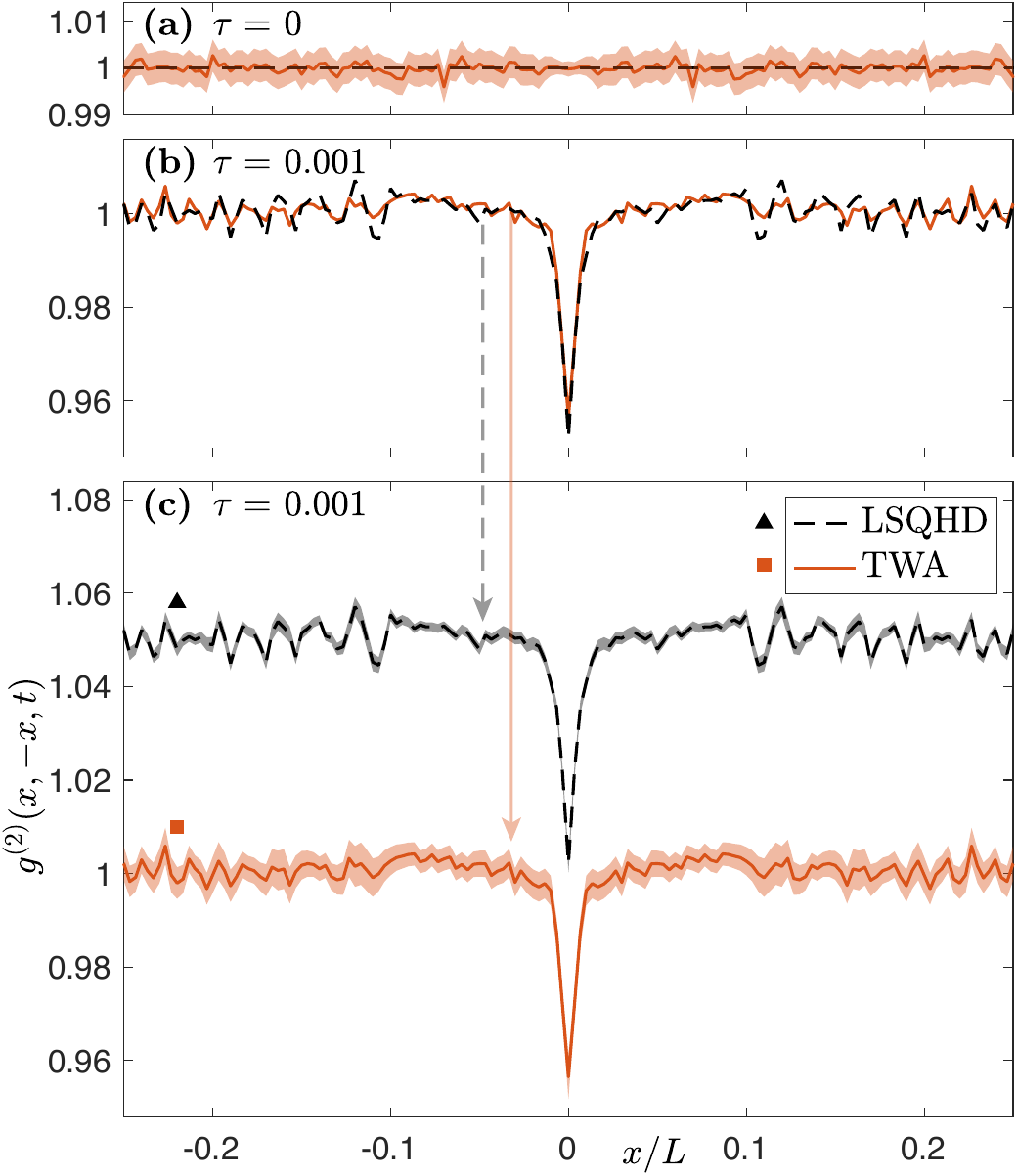}
	\caption{The opposite-point correlation function $g^{(2)}(x,-x,t)$ as in Fig.~\ref{fig:N50_gammabg0.1_sigmabar0.007-g2xmx_comparison}, but for the parameters of Fig.~\ref{fig:N500_gammabg0.01_sigmabar0.007-density_comparison_tend}. To examine the correlations resulting from the density bump itself we plot between $|x|/L=0.25$. Due to the periodic boundary conditions of the system, values outside of this domain correspond to probing the opposite-point correlations of the background. The results are presented similarly to Fig. \ref{fig:N500_gammabg0.01_sigmabar0.007-g2xx_comparison_tend} where the upward shift of the LSQHD curve in (c) is now only 0.05.}
	\label{fig:N500_gammabg0.01_sigmabar0.007-g2xmx_comparison_tend}
\end{figure}

In Figures \ref{fig:N500_gammabg0.01_sigmabar0.007-g2xx_comparison_tend} and \ref{fig:N500_gammabg0.01_sigmabar0.007-g2xmx_comparison_tend} we plot the same- and opposite-point correlation functions respectively. Again, the oscillations which develop coincide with those in the real-space density. However, in this case, the oscillations in $g^{(2)}(x,x,t)$ sit predominantly above the background level, and now oscillate with twice the frequency. Here the predictions of these correlations in the LSQHD (equally, stochastic Bogoliubov) and truncated Wigner approaches agree well with each other, where the linearized schemes slightly overestimate the oscillation amplitudes in comparison to the truncated Wigner result.

\begin{table}[htbp]
	\begin{center}
		\caption{Simulation times for the results presented in Figs. \ref{fig:N500_gammabg0.01_sigmabar0.007-density_comparison_tend}-\ref{fig:N500_gammabg0.01_sigmabar0.007-g2xx_comparison_tend}. For each stochastic approach the results we present here are an average over 500,000 stochastic trajectories; other details are as in Table \ref{table:data_N50_gammabg0.1_sigmabar0.007}.} \label{table:data_N500_gammabg0.01_sigmabar0.007}
		\def\arraystretch{1.3}
		\setlength{\tabcolsep}{18pt}
		\begin{tabular}{cc}
			\hline\hline
			Method & Time to simulate \\
			\hline
			Full $+P$ & 2.3 hours (until $\tau_{{break}}$)\\
			LSQHD & 6.3 hours \\ 
			Stoch. Bogoliubov & 9.2 hours \\ 
			TWA & 3.6 hours \\ 
			\hline\hline
		\end{tabular}
	\end{center}
\end{table}

Table \ref{table:data_N500_gammabg0.01_sigmabar0.007} shows the time taken to simulate this scenario for each approach. Since the full positive-\textit{P} approach breaks down, we show only the simulation time to reach $\tau_{{break}}$. Here again for this example we find that the LSQHD equations are $\sim \!1.5$ faster to simulate than the stochastic Bogoliubov approach. On the other hand, the truncated Wigner approach took a considerably shorter amount of time to simulate than LSQHD. However, we point out that only one complex deterministic equation must be solved in the truncated Wigner approach whereas the LSQHD approach requires the simultaneous integration of two real deterministic equations and two complex stochastic equations. In light of this, it is quite remarkable then that in Table \ref{table:data_N50_gammabg0.1_sigmabar0.007} the LSQHD equations were faster to simulate than the truncated Wigner approach, although we must also remark that no effort has been made to optimize the integration of equations in any of these approaches. Moreover, the truncated Wigner result presented here still contains more than twice as much noise on average as the LSQHD result, and as such would require the simulation of at least four times as many stochastic trajectories to reduce the noise to a similar level.

Lastly, we check the applicability conditions of the approaches used here. The population growth present in the stochastic Bogoliubov scheme reveals an increase in particle number of $\approx11\%$ by the final time of the simulation, which still remains a reasonable application of the undepleted pump approximation. This is supported by the predicted correlation functions which do not depart far from unity, suggesting that the system remains in the regime of small fluctuations where both linearized positive-\textit{P} schemes are valid. The lattice mode occupation of the truncated Wigner approach here is $N/M=500/300\simeq1.7$. Being larger than one, and considering the 1D nature of the system, we consider this a reasonable \textit{a priori} application of the truncated Wigner approach.

%\FloatBarrier

\section{Conclusions}
\label{Conclusions}
In summary, we have derived a set of SQHD equations, using the positive-\textit{P} phase-space approach, that are capable of describing the exact dynamics of trapped Bose gases, and shown that they reduce to other well known approaches under the appropriate approximations. Linearizing such equations allows us to derive the LSQHD scheme which does not suffer from uncontrollable growth of sampling errors at late times and we provide connections between this scheme and Bogoliubov theories. By exploring the density-density correlation functions of quantum shock waves, which develop from a density bump expanding into a non-zero background, we have demonstrated the usefulness and advantages of the LSQHD scheme through a thorough comparison with other dynamical quantum many-body approaches.

%One limitation of the LSQHD scheme however, is that it is not suitable for situations where the mean-field density $\rho_{0}(\boldsymbol{r},t)$ goes to zero because of the division by $\rho_{0}(\boldsymbol{r},t)$ in the quantum pressure terms of Eqs. \eqref{eq:QHD-b}, \eqref{eq:QHD-c}, \eqref{eq:LSQHD-b}, and \eqref{eq:LSQHD-c}. We point out that this limitation is not restricted to the LSQHD scheme, but is inherent in the standard superfluid hydrodynamic approach itself which suffers from the same problem when the mean-field quantum pressure term is kept.
One limitation of the numerical implementation of the LSQHD scheme, however, is that it is not suitable for situations where the mean-field density $\rho_{0}(\boldsymbol{r},t)$ tends to zero, because of the division by $\rho_{0}(\boldsymbol{r},t)$ in the quantum pressure terms of Eqs. \eqref{eq:QHD-b}, \eqref{eq:QHD-c}, \eqref{eq:LSQHD-b}, and \eqref{eq:LSQHD-c}. We point out that this limitation is not restricted to the LSQHD scheme, but is inherent in the standard superfluid hydrodynamic approach itself which suffers from the same problem when the mean-field quantum pressure term is kept \cite{v=0}. Nevertheless, other interesting dynamical scenarios for which the LSQHD scheme would be an ideal and preferred method include: the evolution of a density dip (which sheds gray solitons as it expands into the background \cite{Dutton2001,Kamchatnov2002,Damski2006,Hoefer2008,Hoefer2009,Watson2022}), a combination of density bumps and dips, and density profiles involving step-like functions.

Future outlooks of this work involve going beyond the LSQHD scheme to consider higher-order terms in the expansion like those included in Eqs. \eqref{eq:L2SQH-a}--\eqref{eq:L2SQH-c} and the effect of the many-body quantum pressure term. For example, it would be extremely interesting and insightful to explore possible new constitutive relations in the gradient expansion scheme of conventional hydrodynamics \cite{Grozdanov-Kovtun}, that might follow from the higher-order expansion terms. Another avenue is to find connections with the theory of nonlinear Luttinger liquids \cite{Imambekov_Science,Imambekov_RMP} or more generally with the theories of nonlinear fluctuating hydrodynamics in 1D  \cite{Spohn2016}. In these theories, nonlinear terms are important for describing anomalous transport in 1D classical fluids and alter the universality class of hydrodynamic fluctuations \cite{Spohn2016,Bulchandani_2021}. We further emphasize here that while in phenomenological theories of fluctuating hydrodynamics the noise and dissipation terms are added to Euler-scale hydrodynamics essentially  by hand, the theory of stochastic quantum hydrodynamics derived here is fully microscopic---with the noise terms representing the intrinsic \emph{quantum} fluctuations. Accordingly, exploring its connections to phenomenological theories may provide new insights into the mechanisms behind the emergence of anomalous transport in lower dimensions and the universality classes of associated fluctuations.

\begin{acknowledgments}
The authors acknowledge stimulating discussions with M. J. Davis and D. M. Gangardt. This work was supported by the Australian Research Council Discovery Project Grants DP170101423 and DP190101515. 
\end{acknowledgments}

%\newpage
\appendix

\section{The truncated Wigner approach} \label{appendix:Wigner}
Similarly to the positive-\textit{P} treatment, the truncated Wigner approach is a stochastic phase-space approach which relies on the conversion of the master equation for the quantum density operator to an evolution equation for the Wigner quasiprobability distribution function. For the Wigner distribution function, however, this evolution equation, with a quartic (two-body) interaction Hamiltonian (such as the case here), contains third-order derivative terms. Such higher-than-second-order derivative terms must be truncated in order to render the evolution equation a true Fokker-Planck equation, which can then be equivalently formulated and simulated in terms of stochastic differential equations for complex $c$-fields \cite{Blakie2008,Steel1998}. As such, this approach is only \textit{approximate}, in contrast to the \textit{exact} formulation of the positive-\textit{P} treatment, and care must be taken to ensure that the contribution of the truncated terms is small (usually this is true when the lattice mode occupation of the system is large, \textit{i.e.} the number of particles $N$ is much larger than the number of required modes $M$ such that $N/M\gg1$ \cite{Sinatra2002}). On the other hand, this approach has the advantage that it does not suffer from dynamical noise growth like the positive-\textit{P} treatment does.

In the case of the Hamiltonian \eqref{H} with $s$-wave scattering interactions, the resulting stochastic differential equations actually take the same form as the standard time-dependent GPE. Meanwhile, beyond-mean-field quantum effects are incorporated via the addition of noise into the initial state of the system $\psi_{W}(\mathbf{r},0)$, which is then sampled stochastically, where each realization evolves in time according to the GPE
\begin{align}
	\frac{\partial\psi_{W}}{\partial t}&=\frac{i}{\hbar}\left[\frac{\hbar^{2}}{2m}\laplacian-V_{\text{ext}}(\mathbf{r},t)-g|\psi_{W}|^{2}\right]\psi_{W}.
\end{align}

For the scenarios presented in Section \ref{sec:Density-density correlations in quantum shock waves}, which are carried out in 1D with $g\rightarrow g_{\text{1D}}$, each trajectory is initialized in a coherent-state using \cite{Steel1998}
\begin{align}
	\psi_{W}(x,0) &= \Psi_{0} + \sum_{j=0}^{M} \eta_{j} \phi_{j}(x)
\end{align}
where $\Psi_{0}=\sqrt{N_{\text{bg}}}[1+\beta e^{-x^{2}/2\sigma^2}]/\sqrt{L}$ is the mean-field initial state, and $\eta_{j}$ is a complex Gaussian noise term with zero mean $\langle\eta_{j}\rangle_{stoch.}=0$ and non-zero correlations $\langle\eta_{j}^{*}\eta_{k}\rangle_{stoch.}=\frac{1}{2}\delta_{jk}$. The sum is taken over $M$ modes of a complete basis $\{\phi_{j}(x)\}$, where $M$ is chosen to be sufficiently large such that the results are independent of $M$. Here we choose the discreet position basis, with $\phi_{j}(x)=\delta(x-j\Delta x)/\sqrt{\Delta x}$. 

In the truncated Wigner formalism, we note that stochastic averages correspond to expectation values of symmetrically ordered products of field creation and annihilation operators. As such, the reduced one-body density matrix (which is normally-ordered) is given by
\begin{align}
	G^{(1)}(x,x^{\prime},t) & \equiv\langle\hat{\Psi}^{\dagger}(x,t)\hat{\Psi}(x^{\prime},t)\rangle\nonumber\\
	&=\langle\psi_{W}^{*}(x,t)\psi_{W}(x^{\prime},t)\rangle-\frac{1}{2}\delta_{c}(x,x^{\prime})
\end{align}
where $\frac{1}{2}\int_{-L/2}^{L/2}\delta_{c}(x,x^{\prime})dx=\frac{1}{2}M\delta_{x,x^{\prime}}$ represents the half quantum of vacuum noise per mode $M$ that is included in the Wigner formalism. Here $\delta_{x,x^{\prime}}$ is the Kronecker delta function which gives unity for $x=x^{\prime}$ and zero otherwise. On a computational grid of spacing $\Delta x = L/\mathcal{N}_{x}$, where $\mathcal{N}_{x}$ is the number of grid points, the projected delta function $\delta_{c}(x,x^{\prime})$ is given by $\delta_{c}(x,x^{\prime})=M\delta_{x,x^{\prime}}/(\mathcal{N}_{x}\Delta x)=M\delta_{x,x^{\prime}}/L$ \cite{Blakie2008}. This means that the reduced one-body density matrix can be computed using
\begin{align}
	G^{(1)}(x,x^{\prime},t) & \equiv\langle\hat{\Psi}^{\dagger}(x,t)\hat{\Psi}(x^{\prime},t)\rangle\nonumber\\
	&=\langle\psi_{W}^{*}(x,t)\psi_{W}(x^{\prime},t)\rangle-\frac{M}{2L}\delta_{x,x^{\prime}},
\end{align}
and the particle number density using
\begin{align}
	G^{(1)}(x,x,t) & \equiv\langle\hat{\Psi}^{\dagger}(x,t)\hat{\Psi}(x,t)\rangle\nonumber\\
	&=\langle|\psi_{W}(x,t)|^{2}\rangle-\frac{M}{2L}.
\end{align}
Similarly, the normally ordered density-density correlation function can be computed as \cite{Ng2019}
\begin{align}
	G^{(2)}(x,x^{\prime},t) & \equiv\langle\hat{\Psi}^{\dagger}(x,t)\hat{\Psi}^{\dagger}(x^{\prime},t)\hat{\Psi}(x^{\prime},t)\hat{\Psi}(x,t)\rangle\nonumber \\
	&= \langle|\psi_{W}(x,t)|^{2}|\psi_{W}(x^{\prime},t)|^{2}\rangle\nonumber\\
	&\quad-\frac{M}{2L}(1+\delta_{x,x^{\prime}})\bigg[\langle|\psi_{W}(x,t)|^{2}\rangle\nonumber\\
	&\qquad\qquad\qquad+\left.\langle|\psi_{W}(x^{\prime},t)|^{2}\rangle-\frac{M}{2L}\right],
\end{align}
and then normalized via
\begin{align}
	g^{(2)}(x,x^{\prime},t)&\equiv\frac{G^{(2)}(x,x^{\prime},t)}{G^{(1)}(x,x,t)~G^{(1)}(x^{\prime},x^{\prime},t)}.
\end{align}

\section{Initial states in the positive-\textit{P} approaches} \label{appendix:initial_states}
As mentioned in Section \ref{sec:Comparison with Bogoliubov approaches} of the main text, due to the normally-ordered nature of the positive-\textit{P} approach, an initial coherent state is described simply by not seeding the initial state with any noise.

In the full positive-\textit{P} approach, this amounts to setting $\rho(\mathbf{r},0)=\rho_{0}(\mathbf{r},0)$, $\boldsymbol{v}(\mathbf{r},0)=\boldsymbol{v}_{0}(\mathbf{r},0)$, and $S(\mathbf{r},0)=S_{0}(\mathbf{r},0)$, or equivalently, $\Psi(\mathbf{r},0)=\Psi_{0}(\mathbf{r},0)$ and $\tilde{\Psi}(\mathbf{r},0)=\Psi_{0}^{*}(\mathbf{r},0)$, where $\rho_{0}(\mathbf{r},0)$ is the desired initial mean-field density profile, and $\boldsymbol{v}_{0}(\mathbf{r},0)$ and $S_{0}(\mathbf{r},0)$ are the desired initial mean-field velocity and phase profiles respectively. For the scenarios presented in Section \ref{sec:Density-density correlations in quantum shock waves} these profiles correspond to $\rho_{0}(x,0)=G^{(1)}(x,x,0)=|\Psi_{0}(x,0)|^{2}=N_{\text{bg}}[1+\beta e^{-x^{2}/2\sigma^2}]^{2}/L$, whereas $v_{0}(x,0)=S_{0}(x,0)=0$.

We initialize the linearized positive-\textit{P} approaches in the same way as the full positive-\textit{P} approach; that is, by initializing everything in the mean-field variables as just discussed, and setting $\delta\rho(\mathbf{r},0)=\delta\boldsymbol{v}(\mathbf{r},0)=\delta S(\mathbf{r},0)=0$ [or equivalently, $\delta\psi(\mathbf{r},0)=\delta\tilde{\psi}(\mathbf{r},0)=0$], i.e., assuming that the fluctuating components of the field operators are initially in a vacuum state.

\section{Matrix product state implementation} \label{appendix:MPS}

To implement matrix product state approaches, the Lieb-Liniger Hamiltonian \cite{LiebLiniger1963}
\begin{align}
    \hat{H}_{\text{LL}} &= \int dx~ \hat{\Psi}^{\dagger} \left(-\frac{\hbar^{2}}{2m} \frac{\partial^{2}}{\partial x^{2}} + V(x)\right) \hat{\Psi}\nonumber\\
    &\qquad\qquad\qquad+\frac{g_{\text{1D}}}{2}\int dx~\hat{\Psi}^{\dagger}\hat{\Psi}^{\dagger}\hat{\Psi}\hat{\Psi} \label{eq:Lieb-Liniger H}
\end{align}
is approximated as a Bose-Hubbard model by discretizing in real-space. Infinite matrix product state (iMPS) methods \cite{White1992,McCulloch2007,Schollwock2011} can then be employed to determine the ground and coherent states of the Bose-Hubbard model and to simulate the ensuing dynamics after performing a trap quench. 

The real-space discretization replaces the field operator $\hat{\Psi}(x_{j})$ by $\hat{b}_{j}/\sqrt{\Delta x}$. Here, $\hat{b}_{j}$ is the bosonic creation operator at site $j$ ($j=1,2,\ldots,M,$ for $M=L/\Delta x$ lattice sites), $\Delta x$ is the lattice spacing and $x_{j}=j\Delta x$ \cite{Cazalilla2003,Fleischhauer2005,Fleischhauer2007}. This discretization results in the Bose-Hubbard Hamiltonian
\begin{align}
    \hat{H}_{\text{BH}}=-J\sum_{j} &\left(\hat{b}_{j}^{\dagger} \hat{b}_{j+1} + \hat{b}_{j+1}^{\dagger} \hat{b}_{j}\right) \nonumber\\
    &+ \frac{U}{2} \sum_{j} \hat{b}_{j}^{\dagger2} \hat{b}_{j}^{2} + \sum_{j} V(x_{j}) \hat{b}_{j}^{\dagger} \hat{b}_{j}, \label{eq:Bose-Hubbard H}
\end{align}
where $J=\hbar^{2}/2m\Delta x^{2}$ and $U=g_{\text{1D}}/\Delta x$. This approximation is valid in the continuum limit $\Delta x\rightarrow0$ \cite{Fleischhauer2007,Fleischhauer2010,Schmidt2010} and small average lattice occupancy, $\langle\hat{b}_{j}^{\dagger}\hat{b}_{j}\rangle\ll1$.

The $t = 0$ ground and coherent states are prepared using the infinite Density Matrix Renormalization Group (iDMRG) algorithm, with $M=1000-4000$ sites per unit cell. For the ground state, the iMPS is optimized for the Hamiltonian \eqref{eq:Bose-Hubbard H} using a bond dimension (basis states kept) of $m=70-80$. This results in an energy variance $\sigma_{E}^{2}/J^{2}$ on the order of $10^{-9}$ per site. On the other hand, the coherent state, which is a classical product state, is prepared with $m=1$ and $U=0$. This produces $\sigma_{E}^{2}/J^{2}\sim10^{-13}$ per site. In each case the trapping potential used to prepare the initial state is given by,
\begin{align}
    V(x_{j}) &= \frac{\hbar^{2}\beta}{2m\sigma^{4}} \left(x_{j}^{2} - \sigma^{2}\right) \left[e^{x_{j}^{2}/2\sigma^{2}} + \beta\right]^{-1}\nonumber\\
    &\qquad\qquad\qquad - \frac{g_{\text{1D}}N_{\text{bg}}}{L}\left[1+\beta e^{-x_{j}^{2}/2\sigma^{2}}\right]^{2} \label{eq:V_trap}
\end{align}
which produces exactly the desired density $\rho_{0}(x_{j},0)=N_{\text{bg}}[1+\beta e^{-x_{j}^{2}/2\sigma^{2}}]^{2}/L$ in the mean-field approximation, i.e., as the ground state solution to the GPE \eqref{eq:GPE}.

The time-evolution of the ground and coherent states are executed using the infinite time-evolving block decimation (iTEBD) algorithm \cite{Vidal2007} with the optimized 4th-order decomposition \cite{Barthel2020}. Here, the quench is performed at $t = 0$ by setting $V(x_{j})=0$ and evolving both states with the Hamiltonian \eqref{eq:Bose-Hubbard H}. For the coherent state, the value of $U$ during the dynamics is set to that of the ground state. A time step of 0.2 (in dimensionless units of time $Jt/\hbar$), and a cutoff density matrix eigenvalue of $10^{-10}-10^{-9}$, corresponding to a cutoff singular value on the order of $10^{-5}$, is used throughout this work. This results in a distance $d\equiv1-|F|$ with the $t = 0$ state, after backwards evolution, of $d \approx10^{-7}-10^{-6}$ per site, where $F$ is the fidelity with the initial state.

\section{Observables in the stochastic Bogoliubov formalism} \label{appendix:stoch.bogo.observables}
In the stochastic Bogoliubov approach, the reduced one-body density matrix is given by
\begin{align}
	G^{(1)}(\mathbf{r},\mathbf{r}^{\prime},t) & \equiv\langle\hat{\Psi}^{\dagger}(\mathbf{r},t)\hat{\Psi}(\mathbf{r}^{\prime},t)\rangle\nonumber\\
	&=\langle[\Psi_{0}^{*}(\mathbf{r},t)+\delta\tilde{\psi}(\mathbf{r},t)]\nonumber\\
	&\qquad\qquad\times[\Psi_{0}(\mathbf{r}^{\prime},t)+\delta\psi(\mathbf{r}^{\prime},t)]\rangle\nonumber \\
	&=\Psi_{0}^{*}(\mathbf{r},t)\Psi_{0}(\mathbf{r}^{\prime},t)+\langle\delta\tilde{\psi}(\mathbf{r},t) \delta\psi(\mathbf{r}^{\prime},t)\rangle,
\end{align}
to second order in fluctuating components.

Hence, the particle number density is simply
\begin{align}
	G^{(1)}(\mathbf{r},\mathbf{r},t) & \equiv\langle\hat{\Psi}^{\dagger}(\mathbf{r},t)\hat{\Psi}(\mathbf{r},t)\rangle\nonumber\\
	&=\left|\Psi_{0}(\mathbf{r},t)\right|^{2}+\langle\delta\tilde{\psi}(\mathbf{r},t) \delta\psi(\mathbf{r},t)\rangle,
\end{align}
as given in the main text.

Normalizing the reduced one-body density matrix and maintaining up to second order terms leads to
\begin{align}
	g^{(1)}&(\mathbf{r},\mathbf{r}^{\prime},t)\nonumber\\
	&\equiv\frac{G^{(1)}(\mathbf{r},\mathbf{r}^{\prime},t)}{\sqrt{G^{(1)}(\mathbf{r},\mathbf{r},t)}\sqrt{G^{(1)}(\mathbf{r}^{\prime},\mathbf{r}^{\prime},t)}}\nonumber\\
	&\approx
	\sqrt{\frac{\Psi_{0}^{*}(\mathbf{r},t)\Psi_{0}(\mathbf{r}^{\prime},t)}{\Psi_{0}(\mathbf{r},t) \Psi_{0}^{*}(\mathbf{r}^{\prime},t)}}
	\left(1 + \frac{ \langle\delta \tilde{\psi}(\mathbf{r},t)\delta \psi(\mathbf{r}^{\prime},t)\rangle}{\Psi_{0}^{*}(\mathbf{r},t)\Psi_{0}(\mathbf{r}^{\prime},t)} \right.\nonumber\\
	& \qquad - \left. \frac{1}{2}\frac{\langle\delta \tilde{\psi}(\mathbf{r},t)\delta \psi(\mathbf{r},t)\rangle}{\Psi_{0}^{*}(\mathbf{r},t)\Psi_{0}(\mathbf{r},t)} - \frac{1}{2}\frac{\langle\delta \tilde{\psi}(\mathbf{r}^{\prime},t)\delta \psi(\mathbf{r}^{\prime},t)\rangle}{\Psi_{0}^{*}(\mathbf{r}^{\prime},t)\Psi_{0}(\mathbf{r}^{\prime},t)} \right).
\end{align}

The density-density correlation function in normally ordered form is given by
\begin{align}
	G^{(2)}(\mathbf{r},\mathbf{r}^{\prime},t) & \equiv\langle\hat{\Psi}^{\dagger}(\mathbf{r},t)\hat{\Psi}^{\dagger}(\mathbf{r}^{\prime},t)\hat{\Psi}(\mathbf{r}^{\prime},t)\hat{\Psi}(\mathbf{r},t)\rangle\nonumber \\
	&\approx|\Psi_{0}(\mathbf{r},t)|^{2}|\Psi_{0}(\mathbf{r}^{\prime},t)|^{2}\nonumber\\
	&\quad+|\Psi_{0}(\mathbf{r},t)|^{2}\langle\delta\tilde{\psi}(\mathbf{r}^{\prime},t)\delta\psi(\mathbf{r}^{\prime},t)\rangle\nonumber\\
	&\quad+|\Psi_{0}(\mathbf{r}^{\prime},t)|^{2}\langle\delta\tilde{\psi}(\mathbf{r},t)\delta\psi(\mathbf{r},t)\rangle\nonumber\\
	&\quad+\Psi_{0}(\mathbf{r},t)\Psi_{0}^{*}(\mathbf{r}^{\prime},t)\langle\delta\tilde{\psi}(\mathbf{r},t)\delta\psi(\mathbf{r}^{\prime},t)\rangle\nonumber\\
	&\quad+\Psi_{0}(\mathbf{r}^{\prime},t)\Psi_{0}^{*}(\mathbf{r},t)\langle\delta\tilde{\psi}(\mathbf{r}^{\prime},t)\delta\psi(\mathbf{r},t)\rangle\nonumber\\
	&\quad+\Psi_{0}^{*}(\mathbf{r},t)\Psi_{0}^{*}(\mathbf{r}^{\prime},t)\langle\delta\psi(\mathbf{r},t)\delta\psi(\mathbf{r}^{\prime},t)\rangle\nonumber\\
	&\quad+\Psi_{0}(\mathbf{r},t)\Psi_{0}(\mathbf{r}^{\prime},t)\langle\delta\tilde{\psi}(\mathbf{r},t)\delta\tilde{\psi}(\mathbf{r}^{\prime},t)\rangle
\end{align}
up to second order. Normalizing gives
\begin{align}
	g^{(2)}(\mathbf{r},\mathbf{r}^{\prime},t)&\equiv\frac{G^{(2)}(\mathbf{r},\mathbf{r}^{\prime},t)}{G^{(1)}(\mathbf{r},\mathbf{r},t)~G^{(1)}(\mathbf{r}^{\prime},\mathbf{r}^{\prime},t)} \nonumber\\
	&\approx 1+\frac{\langle\delta\tilde{\psi}(\mathbf{r},t)\delta\psi(\mathbf{r}^{\prime},t)\rangle}{\Psi_{0}^{*}(\mathbf{r},t)\Psi_{0}(\mathbf{r}^{\prime},t)}+\frac{\langle\delta\tilde{\psi}(\mathbf{r}^{\prime},t)\delta\psi(\mathbf{r},t)\rangle}{\Psi_{0}^{*}(\mathbf{r}^{\prime},t)\Psi_{0}(\mathbf{r},t)}\nonumber\\
	&\quad+\frac{\langle\delta\psi(\mathbf{r},t)\delta\psi(\mathbf{r}^{\prime},t)\rangle}{\Psi_{0}(\mathbf{r},t)\Psi_{0}(\mathbf{r}^{\prime},t)}+\frac{\langle\delta\tilde{\psi}(\mathbf{r},t)\delta\tilde{\psi}(\mathbf{r}^{\prime},t)\rangle}{\Psi_{0}^{*}(\mathbf{r},t)\Psi_{0}^{*}(\mathbf{r}^{\prime},t)}\nonumber\\
	&=1+2\Re{\frac{\langle\delta\psi(\mathbf{r},t)\delta\psi(\mathbf{r}^{\prime},t)\rangle}{\Psi_{0}(\mathbf{r},t)\Psi_{0}(\mathbf{r}^{\prime},t)}}\nonumber\\
	&\quad+\frac{\langle\delta\tilde{\psi}(\mathbf{r},t)\delta\psi(\mathbf{r}^{\prime},t)\rangle}{\Psi_{0}^{*}(\mathbf{r},t)\Psi_{0}(\mathbf{r}^{\prime},t)}+\frac{\langle\delta\tilde{\psi}(\mathbf{r}^{\prime},t)\delta\psi(\mathbf{r},t)\rangle}{\Psi_{0}^{*}(\mathbf{r}^{\prime},t)\Psi_{0}(\mathbf{r},t)}
\end{align}
to second order also.

\section{Comparisons between an initial coherent state and the full many-body ground state} \label{appendix:GS-MPS_compare}
Here we make comparisons between the results presented in Section \ref{sec:Example1} of the main text (where each of the approaches are initialized in a coherent state) and those obtained from a matrix product state (MPS) calculation initialized in the full many-body ground state of a trapping potential which reproduces $\rho(x,0)=N_{\text{bg}}[1+\beta e^{-x^{2}/2\sigma^2}]^{2}/L$ in the mean-field density, Eq.~\eqref{eq:V_trap}. From here on we refer to this ground-state MPS calculation as GS-MPS.

\subsection{Density $\rho(x,t)$}
\begin{figure}[tbp]
	\centering
	\includegraphics[width=1.0\linewidth]{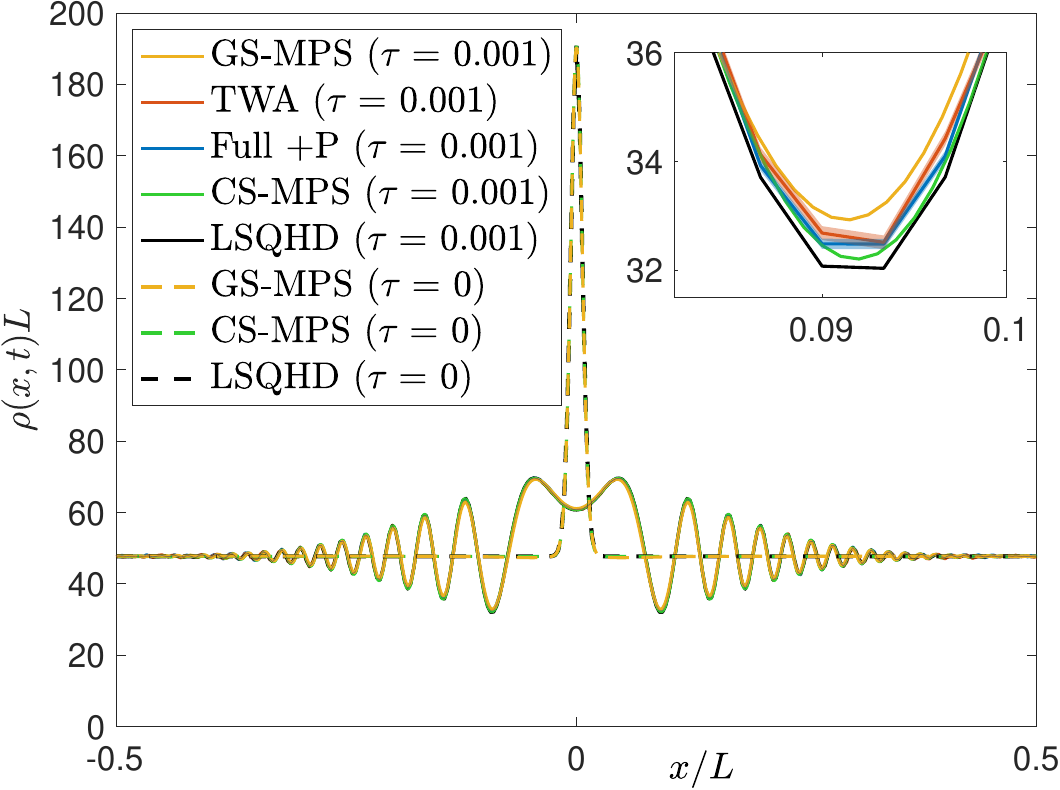}
	\caption{Predictions of the real-space density $\rho(x,t)L$ from different approaches at dimensionless times $\tau=0$ and $\tau=0.001$. %$\tau=t\hbar/mL^{2}$. 
	The inset shows the details of the first density trough. The order of the figure legend coincides with the vertical height order of the results. The shaded regions around the stochastic results denote one standard error of uncertainty, which for LSQHD is on the order of the linewidth used here.}
	\label{fig:N50_gammabg0.1_sigmabar0.007-density_comparison_GS-MPS}
\end{figure}

\begin{figure}[htbp]
	\centering
	\includegraphics[width=1.0\linewidth]{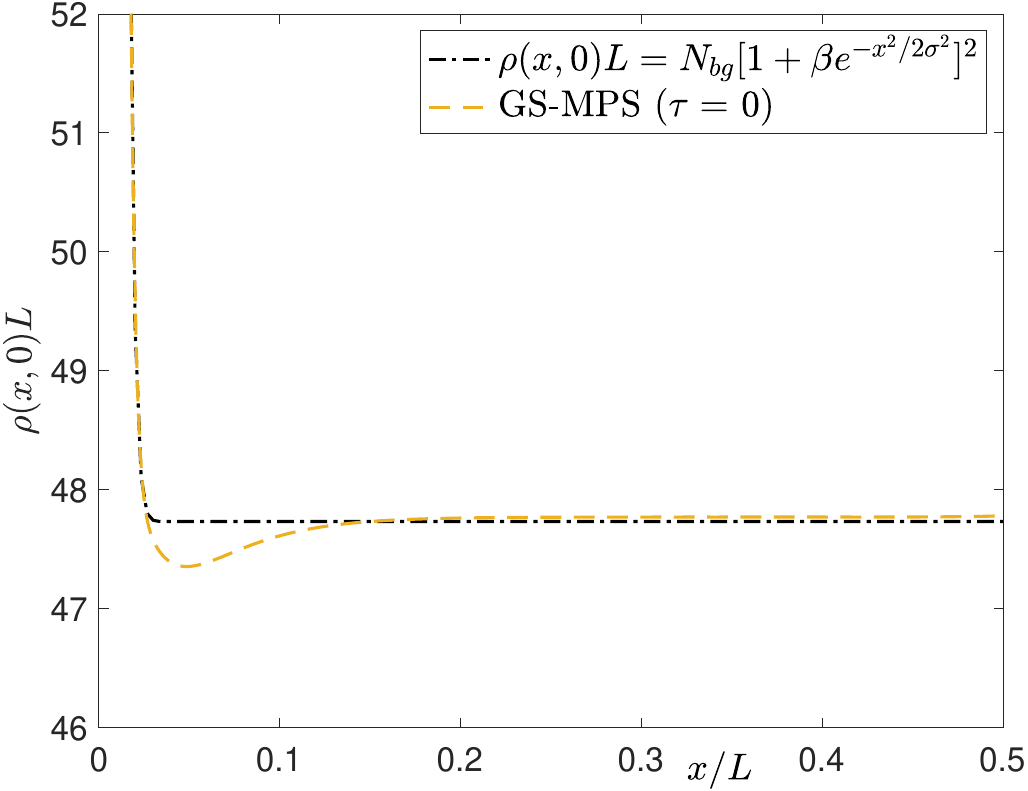}
	\caption{Comparison of the initial real space density $\rho(x,0)L=N_{\text{bg}}[1+\beta e^{-x^{2}/2\sigma^2}]^{2}$ with the GS-MPS result obtained by determining the ground state of the system with Eq.~ \eqref{eq:V_trap} as the trapping potential.}
	\label{fig:N50_gammabg0.1_sigmabar0.007-density_comparison_GS-MPSt0}
\end{figure}

On a global scale, the densities predicted by each method from Sec. \ref{sec:Example1} agree well with the GS-MPS calculation. We do point out however, that the effect of quantum fluctuations can still be discerned by examining the peaks and troughs of the oscillatory wave-train in the time-evolved density $\rho(x,t)$. In Figure \ref{fig:N50_gammabg0.1_sigmabar0.007-density_comparison_GS-MPS} we show the details of the first density trough. Here we see that the full positive-\textit{P}, truncated Wigner, and CS-MPS results agree with each other within one standard error and that the quantum fluctuations present in these approaches slightly diminish the oscillation amplitude as compared with LSQHD (and stochastic Bogoliubov which is not plotted here), whose density is equivalent to the mean-field density at first order in fluctuations. Moreover, the oscillation amplitude is diminished further for the GS-MPS result, where the initial state includes quantum fluctuations already and the initial density profile departs slightly from the coherent state $\rho(x,0)$ since it is found as a many-body ground state of the trapping potential, Eq.~\eqref{eq:V_trap}, which only produces $\rho(x,0)=N_{\text{bg}}[1+\beta e^{-x^{2}/2\sigma^2}]^{2}/L$ in the mean-field. This difference in initial density profiles is shown in Fig. \ref{fig:N50_gammabg0.1_sigmabar0.007-density_comparison_GS-MPSt0}.

\subsection{Correlation functions $g^{(2)}(x,x,t)$ and $g^{(2)}(x,-x,t)$ from the MPS approaches}

\begin{figure}[htbp]
	\centering
	\includegraphics[width=1.0\linewidth]{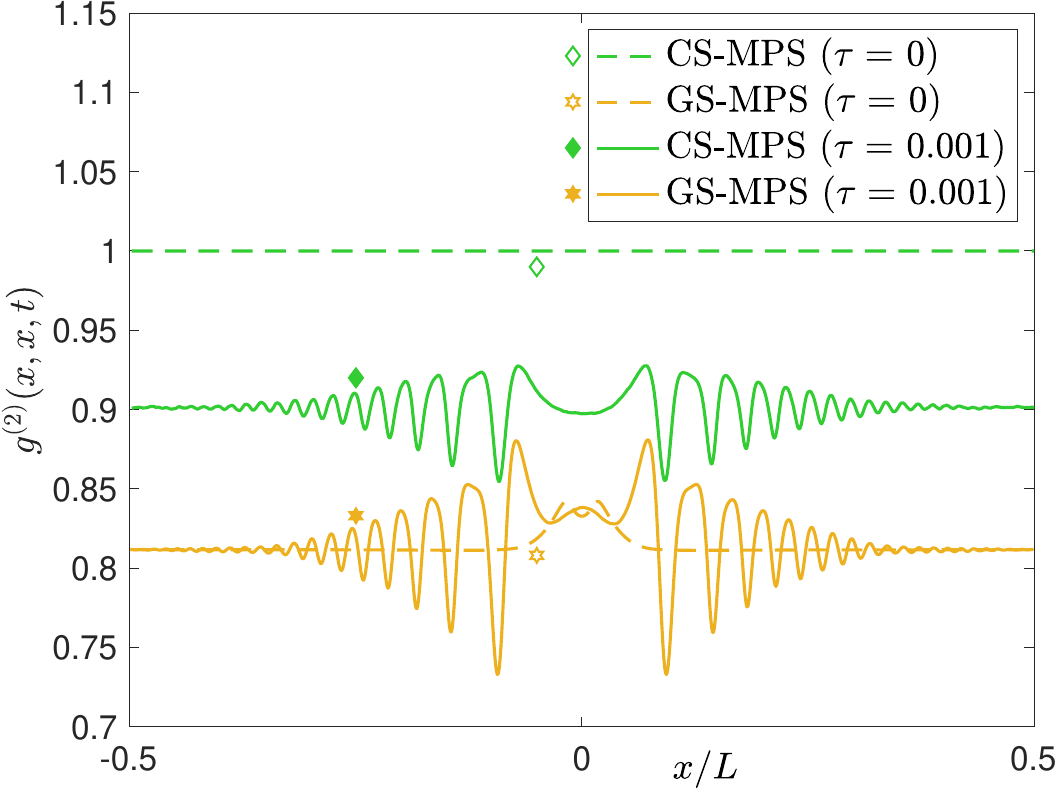}
	\caption{The same-point correlation function $g^{(2)}(x,x,t)$ of Fig. \ref{fig:N50_gammabg0.1_sigmabar0.007-g2xx_comparison}, as predicted by CS-MPS and GS-MPS at dimensionless times $\tau=0$ and $\tau=0.001$. The results are shown as per the figure legend and associated marker symbols.}
	\label{fig:N50_gammabg0.1_sigmabar0.007-MPSg2xx_comparison}
\end{figure}

\begin{figure}[htbp]
	\centering
	\includegraphics[width=1.0\linewidth]{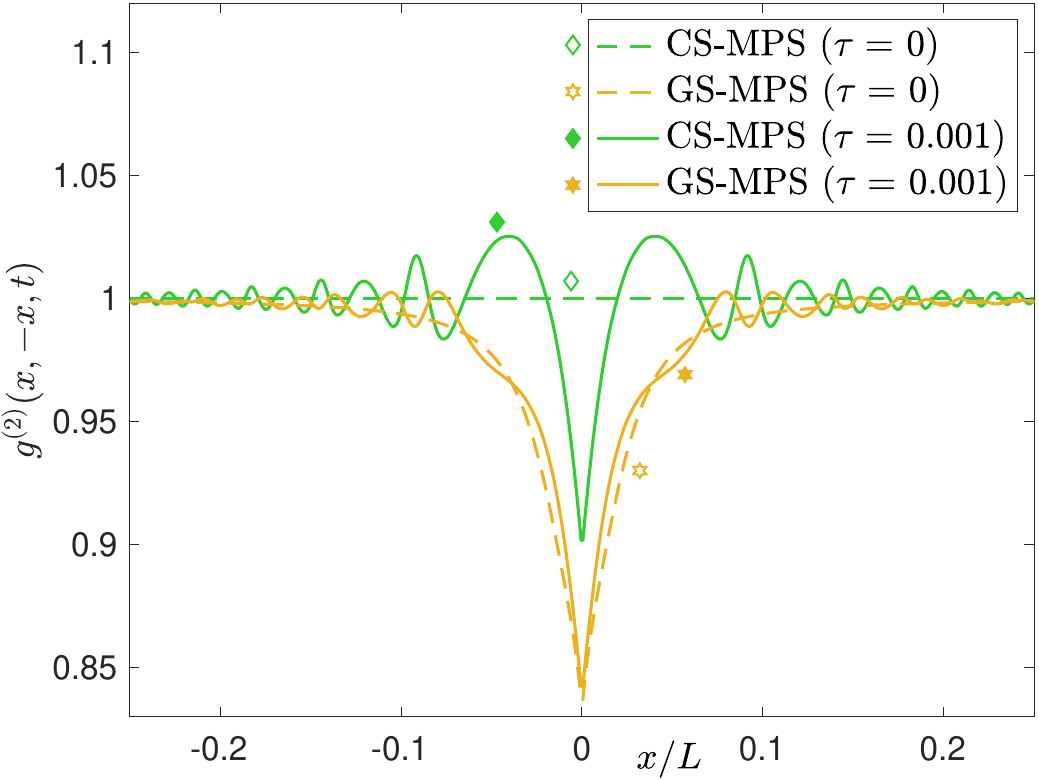}
	\caption{The opposite-point correlation function $g^{(2)}(x,-x,t)$ of Fig. \ref{fig:N50_gammabg0.1_sigmabar0.007-g2xmx_comparison}, as predicted by CS-MPS and GS-MPS at dimensionless times $\tau=0$ and $\tau=0.001$. The results are shown as per the figure legend and associated marker symbols.}
	\label{fig:N50_gammabg0.1_sigmabar0.007-MPSg2xmx_comparison}
\end{figure}

In figures \ref{fig:N50_gammabg0.1_sigmabar0.007-MPSg2xx_comparison} and \ref{fig:N50_gammabg0.1_sigmabar0.007-MPSg2xmx_comparison} we compare the differences in predictions of $g^{(2)}(x,x,t)$ and $g^{(2)}(x,-x,t)$, respectively, between an initial coherent state and the full many-body ground state. For clarity we compare only the CS-MPS result with the GS-MPS calculation and we recall the good agreement of each stochastic approach with the CS-MPS result, especially for LSQHD.

We first point out the good qualitative agreement between the two different initializations; the coherent-state simulations correctly capture all of the broad features present in the exact ground-state calculation for both $g^{(2)}(x,x,t)$ and $g^{(2)}(x,-x,t)$.

The primary quantitative disagreement for $g^{(2)}(x,x,t)$ is the shift in background value due to the initial correlations present in the GS-MPS result. Additionally, the oscillation amplitude is reduced in the CS-MPS result compared with GS-MPS. For $g^{(2)}(x,-x,t)$ the primary differences between the coherent state and ground state initializations are the opposite orientation of the oscillations, and this time their larger amplitude in the CS-MPS result compared to GS-MPS.

\section{Comparison of the density and correlation function $g^{(1)}(\mathbf{r},\mathbf{r}^{\prime},t)$ between the LSQHD and stochastic Bogoliubov approaches} \label{appendix:g1_compare}

\subsection{Density comparisons} \label{appendix:g1_compare_density}
\begin{figure}[htbp]
	\includegraphics[width=1.0\linewidth]{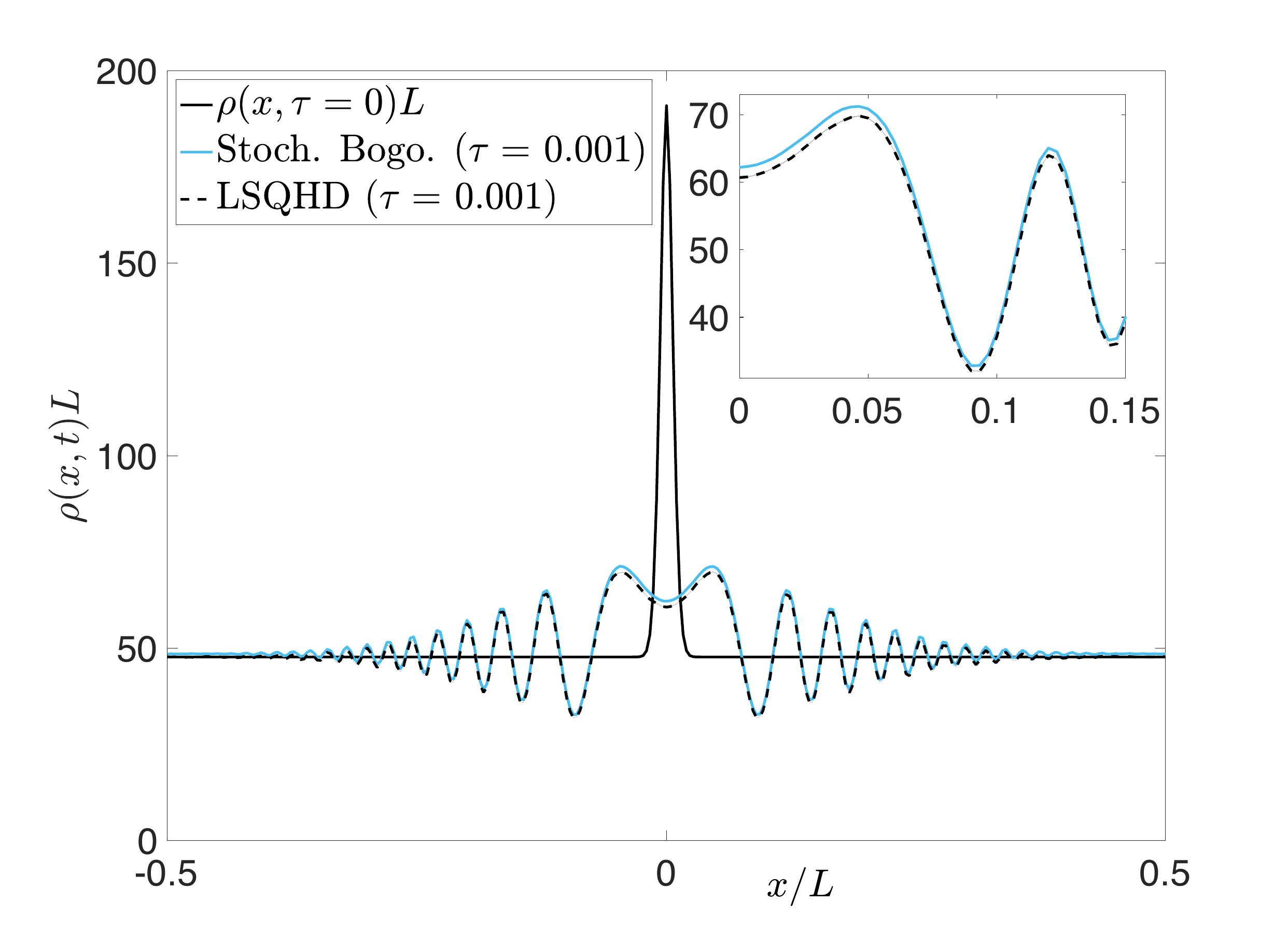}
	\caption{The real-space density $\rho(x,t)$ for the scenario presented in Fig. \ref{fig:N50_gammabg0.1_sigmabar0.007-density_comparison}; here we compare the results from the LSQHD and stochastic Bogoliubov (Stoch. Bogo.) approaches.}
	\label{fig:N50_gammabg0.1_sigma0.007_lat300_density_comparisonSB}
\end{figure}

\begin{figure}[htbp]
	\includegraphics[width=1.0\linewidth]{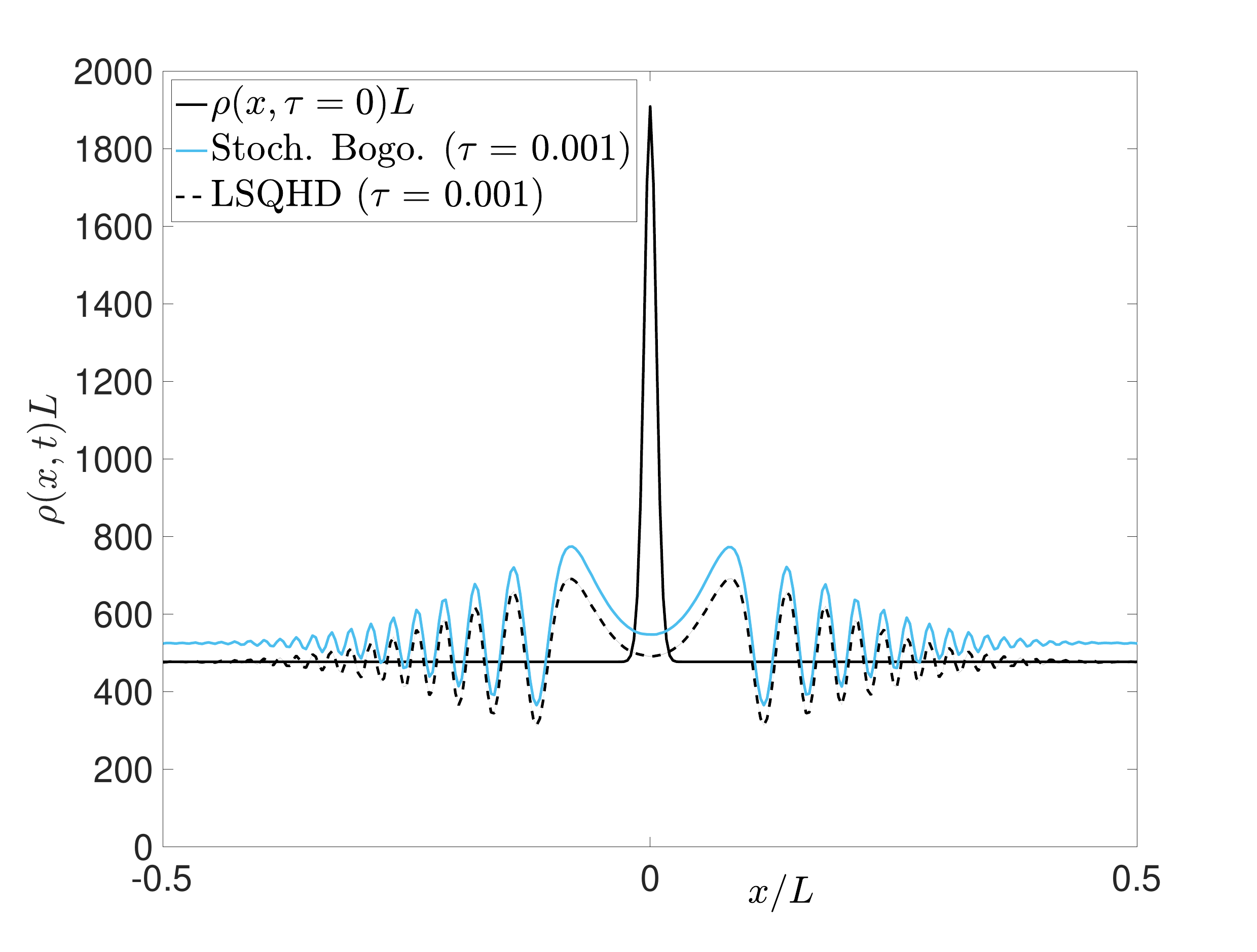}
	\caption{The real-space density $\rho(x,t)$ for the scenario presented in Fig.~ \ref{fig:N500_gammabg0.01_sigmabar0.007-density_comparison_tend}; here we compare the results from the LSQHD and stochastic Bogoliubov (Stoch. Bogo.) approaches.}
	\label{fig:N500_gammabg0.01_sigma0.007_lat300_density_comparisonSB}
\end{figure}

A comparison of the density predictions of the LSQHD and stochastic Bogoliubov approaches for the two examples given in the main text are provided in Figures \ref{fig:N50_gammabg0.1_sigma0.007_lat300_density_comparisonSB} and \ref{fig:N500_gammabg0.01_sigma0.007_lat300_density_comparisonSB}. The LSQHD prediction is given by equation \eqref{eq:G1rr_LSQH}, which differs from the stochastic Bogoliubov expression \eqref{eq:G1rr_SB} by the term responsible for particle number growth, $\langle\delta\tilde{\psi}(\mathbf{r},t) \delta\psi(\mathbf{r},t)\rangle$. This particle number growth can be observed in the figures as the discrepancy between the respective curves at dimensionless time $\tau=0.001$.

\subsection{Correlation function $g^{(1)}(\mathbf{r},\mathbf{r}^{\prime},t)$} \label{appendix:g1_compare_correlations}
To identify the difference between the LSQHD expression for $g^{(1)}(\mathbf{r},\mathbf{r}^{\prime})$ \eqref{eq:g1_LSQH} and the stochastic Bogoliubov expression \eqref{eq:g1_SB}, one can apply Madelung's transformation $\Psi_{0}(\mathbf{r},t)=\sqrt{\rho_{0}(\mathbf{r},t)}e^{iS_{0}(\mathbf{r},t)}$ and substitute \eqref{eq:g1_normal} into the stochastic Bogoliubov result \eqref{eq:g1_SB}. Doing so leads to
\begin{align}
	g&^{(1)}(\mathbf{r},\mathbf{r}^{\prime})\nonumber\\
	&\approx e^{-i\left[S_{0}(\mathbf{r}) - S_{0}(\mathbf{r}^{\prime})\right]} \nonumber\\
	&\quad\times \left( 1+ \expval{\delta S(\mathbf{r})\delta S(\mathbf{r}^{\prime})} - \frac{1}{2} \left[\expval{\delta S^{2}(\mathbf{r})} + \expval{\delta S^{2}(\mathbf{r}^{\prime})}\right] \right. \nonumber\\
	& \qquad\quad + \frac{\expval{\delta \rho(\mathbf{r})\delta \rho(\mathbf{r}^{\prime})}}{4\rho_{0}(\mathbf{r})\rho_{0}(\mathbf{r}^{\prime})} - \frac{1}{8} \left[\frac{\expval{\delta\rho^{2}(\mathbf{r})}}{\rho_{0}^{2}(\mathbf{r})} + \frac{\expval{\delta\rho^{2}(\mathbf{r}^{\prime})}}{\rho_{0}^{2}(\mathbf{r}^{\prime})}\right] \nonumber\\
	&\qquad\qquad\quad + \left. \frac{i}{2} \left[\frac{\expval{\delta \rho(\mathbf{r})\delta S(\mathbf{r}^{\prime})}}{\rho_{0}(\mathbf{r})} -  \frac{\expval{\delta S(\mathbf{r})\delta \rho(\mathbf{r}^{\prime})}}{\rho_{0}(\mathbf{r}^{\prime})}\right] \right), \label{eq:g1_SB-hydro_variables}
\end{align}
which differs from the LSQHD expression \eqref{eq:g1_LSQH} in that the term
\begin{align}
	\frac{i}{2} \left[\frac{\expval{\delta \rho(\mathbf{r}^{\prime})\delta S(\mathbf{r}^{\prime})}}{\rho_{0}(\mathbf{r}^{\prime})} - \frac{\expval{\delta S(\mathbf{r})\delta \rho(\mathbf{r})}}{\rho_{0}(\mathbf{r})}\right] \label{eq:g1_SB_missing-term}
\end{align}
is absent here in Eq. \eqref{eq:g1_SB-hydro_variables}.

\begin{figure}[tbp]
	\includegraphics[width=1.0\linewidth]{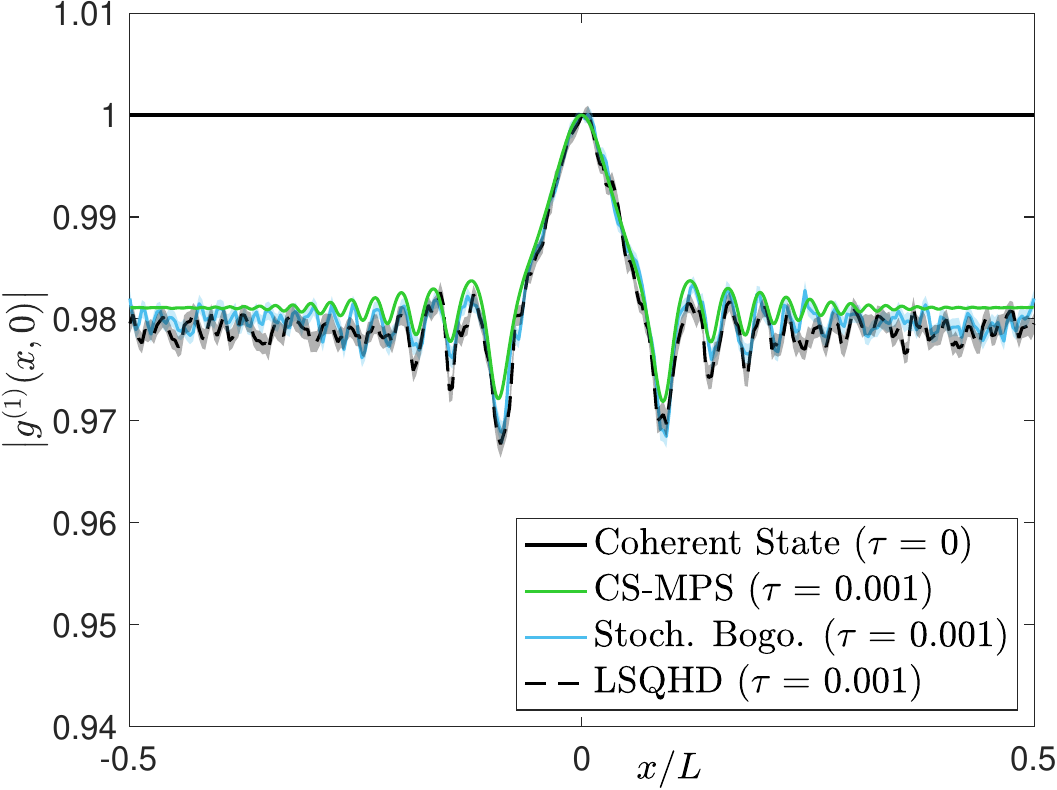}
	\caption{The correlation function $g^{(1)}(x,0,t)$ for the scenario presented in Figure \ref{fig:N50_gammabg0.1_sigmabar0.007-density_comparison} of the main text. Here however, the stochastic results are an average over only 500,000 trajectories and the shaded regions denote one standard error of uncertainty. The legend order resembles the position of each result; the CS-MPS result is smooth and sits slightly above both the solid blue (light gray) Stoch. Bogo. result and the dashed black LSQHD result, which themselves lie almost directly on top of each other and are difficult to distinguish.}
	\label{fig:N50_gammabg0.1_sigma0.007_lat300_coherent_g1x0}
\end{figure}

\begin{figure}[htbp]
	\includegraphics[width=1.0\linewidth]{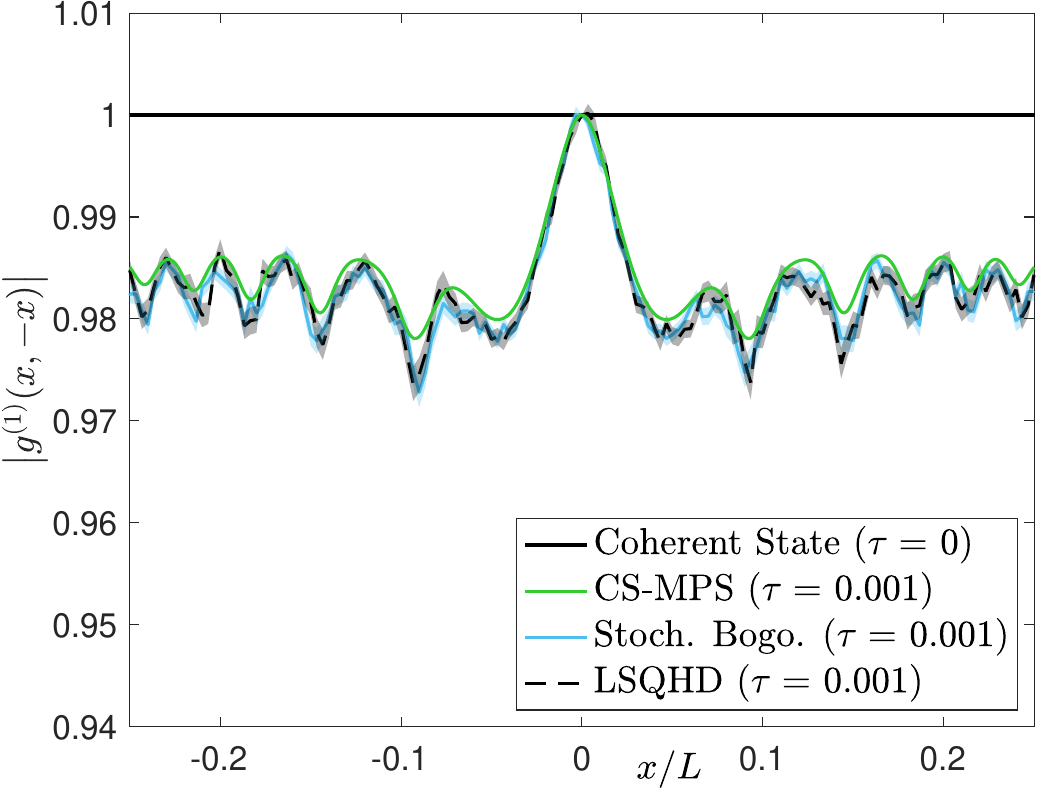}
	\caption{The correlation function $g^{(1)}(x,-x,t)$ for the scenario presented in Figure \ref{fig:N50_gammabg0.1_sigmabar0.007-density_comparison} of the main text. Here however, the stochastic results are an average over only 500,000 trajectories and the shaded regions denote one standard error of uncertainty. The legend order resembles the position of each result; the CS-MPS result is smooth and sits slightly above both the solid blue (light gray) Stoch. Bogo. result and the dashed black LSQHD result, which themselves lie almost directly on top of each other and are difficult to distinguish.}
	\label{fig:N50_gammabg0.1_sigma0.007_lat300_coherent_g1xmx}
\end{figure}

We explore the effect of this term by numerically computing $g^{(1)}(x,x^{\prime},t)$ in the LSQHD, stochastic Bogoliubov, and CS-MPS approaches for the scenario of Fig. \ref{fig:N50_gammabg0.1_sigmabar0.007-density_comparison} given in Sec. \ref{sec:Example1} of the main text. A comparison of the slices $g^{(1)}(x,0,t)$ and $g^{(1)}(x,-x,t)$ are shown in Figs. \ref{fig:N50_gammabg0.1_sigma0.007_lat300_coherent_g1x0} and \ref{fig:N50_gammabg0.1_sigma0.007_lat300_coherent_g1xmx}, respectively. The differences between the predictions of the LSQHD and stochastic Bogoliubov approaches is on the order of the uncertainty in these calculations, and hence we conclude that the term \eqref{eq:g1_SB_missing-term} responsible for the difference in expressions \eqref{eq:g1_SB-hydro_variables} and \eqref{eq:g1_LSQH} is negligible here.

Further exploration reveals that the other term containing cross products of $\delta \rho$ and $\delta S$, 
\begin{align}
    \frac{i}{2} \left[\frac{\expval{\delta \rho(\mathbf{r})\delta S(\mathbf{r}^{\prime})}}{\rho_{0}(\mathbf{r})} -  \frac{\expval{\delta S(\mathbf{r})\delta \rho(\mathbf{r}^{\prime})}}{\rho_{0}(\mathbf{r}^{\prime})}\right],
\end{align}
also contributes negligibly to $g^{(1)}(x,x^{\prime},t)$ in this scenario.

Additionally, the LSQHD and stochastic Bogoliubov results lie close to the numerically exact CS-MPS prediction, confirming the validity and accuracy of these linearized schemes in computing $g^{(1)}(x,x^{\prime},t)$ for this scenario.

\FloatBarrier
%\bibliography{Refs_SQHD}

\begin{thebibliography}{112}
\bibitem [{Ito()}]{Ito_vs_Stratonovich}%
	\BibitemOpen
	\href@noop {} {}\bibinfo {howpublished} {Whilst the stochastic Bogoliubov
		equations are often written in \^{I}to form, we remind the reader that the
		\^{I}to and Stratonovich forms of equations \eqref{eq:pos-P-1} and
		\eqref{eq:pos-P-2} are equivalent (up to a physically irrelevant constant
		phase shift). Hence, the \^{I}to and Stratonovich forms of linearized schemes
		based on these equations are also equivalent.}\BibitemShut {Stop}%
\bibitem [{v=0()}]{v=0}%
	\BibitemOpen
	\href@noop {} {}\bibinfo {howpublished} {Moreover, in situations where the
		density goes precisely to zero, the velocity field is no longer well-defined,
		and even the classical hydrodynamic equations \eqref{eq:CHD-a} and
		\eqref{eq:CHD-b} are no longer suitable to describe the system in that
		case.}\BibitemShut {Stop}%
\end{thebibliography}


\end{document}